\documentclass[12pt, draftclsnofoot, onecolumn]{IEEEtran}
\usepackage[super]{nth}
\usepackage{amsthm}

\usepackage{cite}
\usepackage{amsmath,amssymb,amsfonts}
\usepackage{graphicx}
\usepackage{fancyhdr}
\usepackage{textcomp}
\usepackage{xcolor}
\usepackage{algorithmicx,algpseudocode}
\usepackage{algpseudocode,algorithm}
\usepackage{booktabs}
\usepackage{kotex}

\algrenewcommand\algorithmicrequire{\textbf{Precondition:}}
\algrenewcommand\algorithmicensure{\textbf{Postcondition:}}

\theoremstyle{definition}

\newtheorem{definition}{Definition}

\usepackage{caption}
\usepackage{subcaption}

\begin{document}
	
	\title{Joint Distributed Link Scheduling and Power Allocation for Content Delivery in Wireless Caching Networks
	}
	
	\author{Minseok Choi,~\IEEEmembership{Member,~IEEE,} Andreas F. Molisch,~\IEEEmembership{Fellow,~IEEE}, and Joongheon Kim,~\IEEEmembership{Senior Member,~IEEE}
		\thanks{
			The work was financially supported in part by NSF under projects NSF CCF-1423140 and NSF CNS-1816699, and IITP grant funded by the Korea government (MSIP) (No. 2017-0-00068, A Development of Driving Decision Engine for Autonomous Driving using Driving Experience Information)
		}
		\thanks{M. Choi and Andreas F. Molisch are with the Department of Electrical and Computer Engineering, University of Southern California, Los Angeles, USA, e-mail: choimins@usc.edu, molisch@usc.edu.}
		\thanks{J. Kim is with the School of Electrical Engineering, Korea University, Seoul 02841, South Korea, e-mail: joongheon@korea.ac.kr}
		\thanks{J. Kim is a corresponding author of this paper.}
	}
	
	\maketitle
	
	\begin{abstract}
		
		In wireless caching networks, the design of the content delivery method must consider random user requests, caching states, network topology, and interference management.
		In this paper, we establish a general framework for content delivery in wireless caching networks without stringent assumptions that restrict the network structure, delivery link, and interference model.
		Based on the framework, we propose a dynamic and distributed link scheduling and power allocation scheme for content delivery that is assisted by belief-propagation (BP) algorithms. 
		Considering content-requesting users and potential caching nodes, 
		the scheme achieves three critical purposes of wireless caching networks: 1) limiting the delay of user request satisfactions, 2) maintaining the power efficiency of caching nodes, and 3) managing interference among users. 
		In addition, we address the intrinsic problem of the BP algorithm in our network model, proposing a matching algorithm for one-to-one link scheduling. 
		Simulation results show that the proposed scheme provides almost the same delay performance as the optimal scheme found through an exhaustive search at the expense of a little additional power consumption and does not require a clustering method and orthogonal resources in a large-scale D2D network.
		
	\end{abstract}
	
	
 	\section{Introduction}
	\label{sec:intro}
	
	Multimedia services, in particular, on-demand video streaming, accounts for a large portion of the global wireless data traffic \cite{cisco}. 
	In such services, a relatively small number of popular contents is requested multiple times by multiple users. 
	This has given rise to the idea of supporting overlapped user demands for popular contents by storing them on independent entities close to the end user, such as femto-base stations (BSs) \cite{femtocaching,CM2014Bastug,CM2014Wang,CommMag2013Golrezaei}, or on user devices \cite{CommMag2013Golrezaei,JSAC2016Ji,TIT2016Ji}, during off-peak hours.
	Such a wireless caching technique can reduce or eliminate redundant backhaul communications stemming from overlapped content requests.
	In addition to throughput, which has been widely used as a key performance criterion for video services \cite{TMM2009He}, the delay of the delivery is also an important measure and can be improved through wireless edge caching.
	
	There are two main stages in wireless caching networks: content caching and content delivery. 
	Various probabilistic caching policies for stochastic wireless networks were proposed to maximize the successful delivery rate \cite{TWC2016Chae}, minimize a generic cost function \cite{JSAC2016Gregori}, or serve consecutive user demands \cite{ICC2019Choi}.
	Content caching distributions have been shown to be relatively robust; i.e., deviations of the actual from the optimum caching distribution results in small performance losses \cite{JSAC2016Ji}. 
	We thus assume in this paper that the caching distribution is provided. 
	
	Delivery of cached content is fundamentally different from wireless peer-to-peer (P2P) communication: whereas in P2P scenarios, unique pairs of transmitters and receivers want to communicate with each other, as described in \cite{JSAC2012Rangan}, in cached content delivery there may exist multiple caching nodes that can provide a file to the requesting user. 
	Specifically, wireless caching networks have the distinct characteristics that users are able to receive contents from any caching node that can deliver the content successfully, and delivery can be scheduled sporadically provided that their requests are filled well within the delay time limit.
	Similarly, caching nodes can transmit the content to any user requesting one of their cached contents.
	Thus, node association, i.e., the selection of the best source node for the user, becomes an important aspect of the delivery process in wireless caching networks. 
	
	In much of the information-theoretic literature on video caching \cite{JSAC2016Ji,TIT2015Ji,TIT2017Jeon}, node association has not been considered an issue, because the derivations assume a clustering structure, such that communication is possible for only one delivery link in each cluster at a time, and -- even if multiple nodes can act as sources nodes -- it is unimportant which source node is selected. 
	This model, although useful for closed-form derivations, is not, however, practical. 
	Although users and helpers may be distributed under a physical channel model, in \cite{TWC2017Chen} and \cite{TWC2018Amer} cooperative transmission schemes were proposed with the assumption that only one delivery link can be active in a cluster to avoid interference.
	When node association is considered, the traditional method is to select the node that has the strongest channel condition \cite{TWC2016Chae}.
	Meanwhile, for cases where the same content with different qualities (and thus different file sizes) is stored on different nodes, the authors of \cite{JSAC2018Choi,TWC2019Choi} proposed dynamic node association to meet the differentiated quality requirements of users.
	
	Even with the cellular structure where content-requesting users can receive the desired file from the BS to which the user belongs, in many of the existing studies on wireless caching a hard constraint was applied to interference models.
	The scheme presented in \cite{TCOM2018Z_Yang} allows multiple users to request contents in the same picocell by using orthogonal resources, and the authors of this paper proposed a bandwidth allocation rule.
	Joint downlink scheduling and delivery in HetNets were addressed in \cite{TCOM2019Lv}, where the interference effect on a BS from other BSs assumed to transmit with peak power was considered.
	In \cite{TCOM2018J_Yang}, interference effects between a macro BS and femto BSs were captured by using a monetary cost model. 
	However, in \cite{TCOM2019Lv,TCOM2018J_Yang} it was still assumed that no interference exists among users in the same femtocell. 
	Meanwhile, BS association in caching networks was proposed in \cite{Access2019Jing,TCOM2019Kwak} and the interference caused by multiple BSs was considered without any strict constraint for interference models;
	however, adaptive power allocations for interference management were not considered in these studies. 
	
	Device-to-device (D2D) link scheduling with power control in the presence of interference in wireless caching networks was investigated in \cite{TCOM2016Zhang}. 
	The scheme proposed in \cite{TCOM2016Zhang} is based on centralized decisions; therefore, comprehensive knowledge of the entire network is required. 
	A game-theoretic framework for D2D link scheduling in caching networks by balancing network costs and social satisfactions was proposed in \cite{TWC2018La}. 
	In this framework, interference effects are reflected in the power consumption required to satisfy the given target performance.
	In \cite{ICC2019Chuan}, belief propagation (BP) based content delivery in wireless caching networks was addressed; however, adaptive control of the transmit power was not considered.
	
	Because transmit power adjustments significantly assist interference management, the scheme presented in this paper jointly optimizes link scheduling and power allocation according to the stochastic network states in a distributed manner.
	The BP-based scheduling and power control scheme was proposed in the conference version of this paper; however, the current paper presents improvements: 1) It addresses the intrinsic problem of the BP algorithm is addressed for one-to-one link scheduling, and a matching algorithm to resolve this problem is proposed; 2) the model includes D2D-assisted caching networks, and potential helpers can be in the idle state, because their number is greater than that of active users; and 3) extensive simulations show that clustering is not required and compare our scheme with the existing delivery schemes.
	The main contributions are as follows. 
	\begin{itemize}
		\item This paper establishes a general framework for content delivery in wireless caching networks without any stringent assumption that restricts the network structure, delivery link, and interference model.
		In contrast to the existing schemes, the proposed delivery scheme takes into account interference effects from any active caching node; therefore, it does not require clustering with different bandwidth allocations or a \textit{protocol model}, which is limited in terms of supporting many active delivery links simultaneously.
		Based on the framework, in this paper a link scheduling and power allocation policy is proposed that is dynamically adjustable according to the stochastic network states and random user requests.
		The proposed general framework and delivery scheme can be applied to caching helper networks, as well as to D2D-assisted caching networks.
		
		\item The central unit may not be able to control content delivery easily in wireless caching networks because of a lack of knowledge of the entire time-varying network.
		We thus propose a BP-based algorithm that facilitates distributed decisions on link scheduling and power allocation at every caching node.
		The proposed BP and matching algorithms can be implemented at each caching node separately by exchanging local information with near devices only. 
		The distributed delivery scheme is more applicable than centralized control to the practical scenario where many devices are involved, and thus, it is difficult to centrally gather network information, e.g., channel gains and user queue states. 
		
		\item We address the essential problem of the BP algorithm that, although one-helper-to-one-user scheduling is considered in the problem formulation, the result of the BP algorithm occasionally suggests many-to-one scheduling.
		Therefore, to tackle this problem we propose a matching algorithm that is applied after the BP algorithm. 
		The proposed matching for one-to-one link scheduling is based on the well-known deferred acceptance (DA) algorithm \cite{Book:matching}; however, its application for improving the BP results is non-trivial.
		
		\item We conduct simulations to evaluate the performance of the proposed link scheduling and power allocation policy. 
		In a helper network, it is shown that the proposed scheme provides limits on the averaged queueing delay that are very similar to those obtained with the exhaustive search technique, which is the optimal centralized decision mechanism.
		This good performance is achieved at the expense of power consumption, which is increased by 70\%. 
		In a large-scale D2D network, the proposed scheme has the potential to support more than 200\% of the request rate supported by clustering methods. 
		
	\end{itemize}
	
	The rest of the paper is organized as follows. 
	The wireless caching network is described in Section \ref{sec:system_model}. 
	The joint optimization problem of link scheduling and power allocation is formulated in Section \ref{sec:joint_scheduling}.
	The proposed BP-based content delivery policy and the matching algorithm for one-to-one link scheduling are presented in Sections \ref{sec:BP} and \ref{sec:matching}, respectively.
	Our simulation results are presented in Section \ref{sec:numerical_results}, and Section \ref{sec:conclusion} concludes this paper.
	
	\section{System Model}
	\label{sec:system_model}

	\begin{figure}[t]
		\minipage{0.36\textwidth}
		\includegraphics[width=\linewidth]{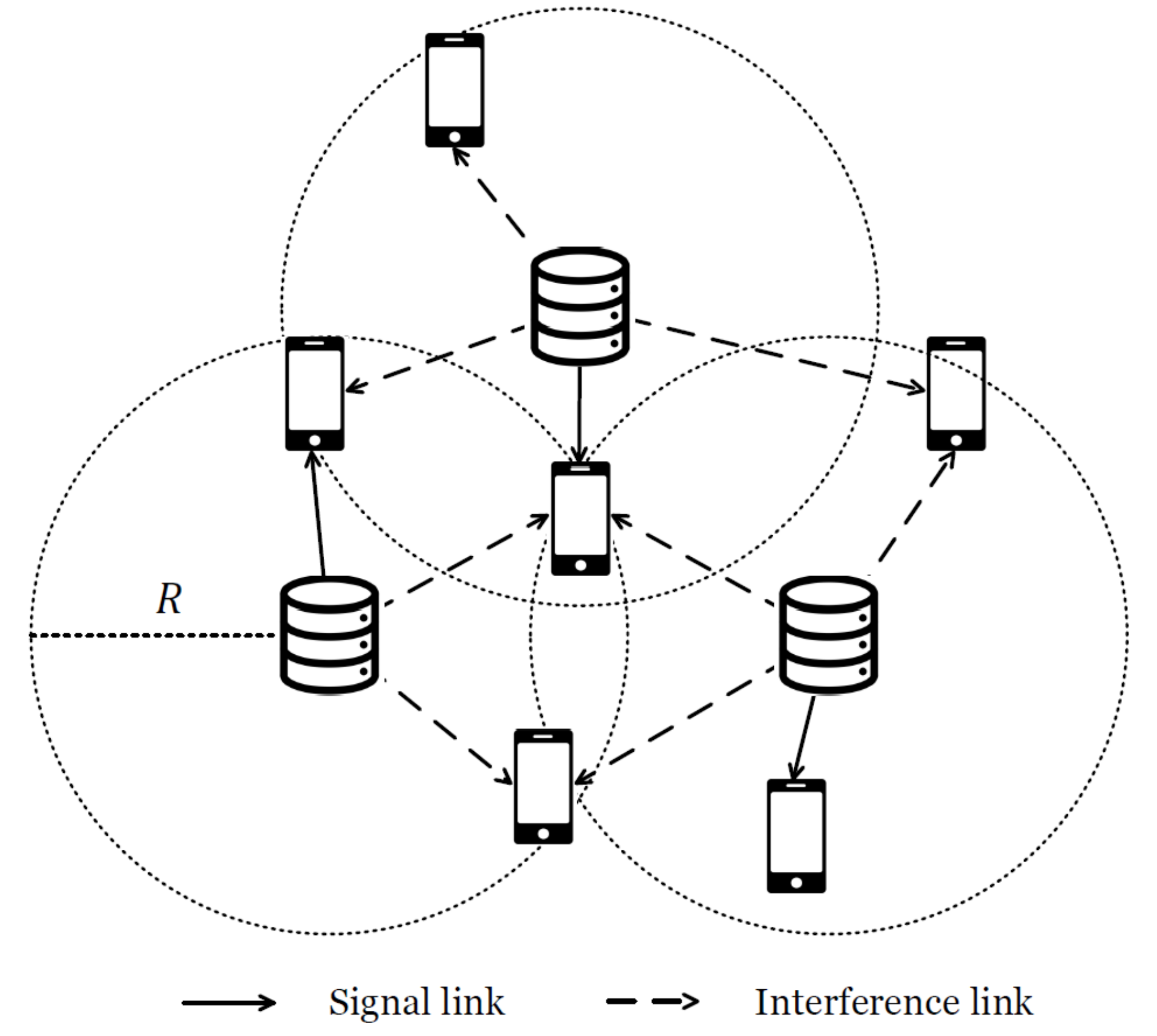}
		\caption{Caching helper network}
		\label{fig:network_model}
		\endminipage\hfill
		\minipage{0.33\textwidth}
		\includegraphics[width=\linewidth]{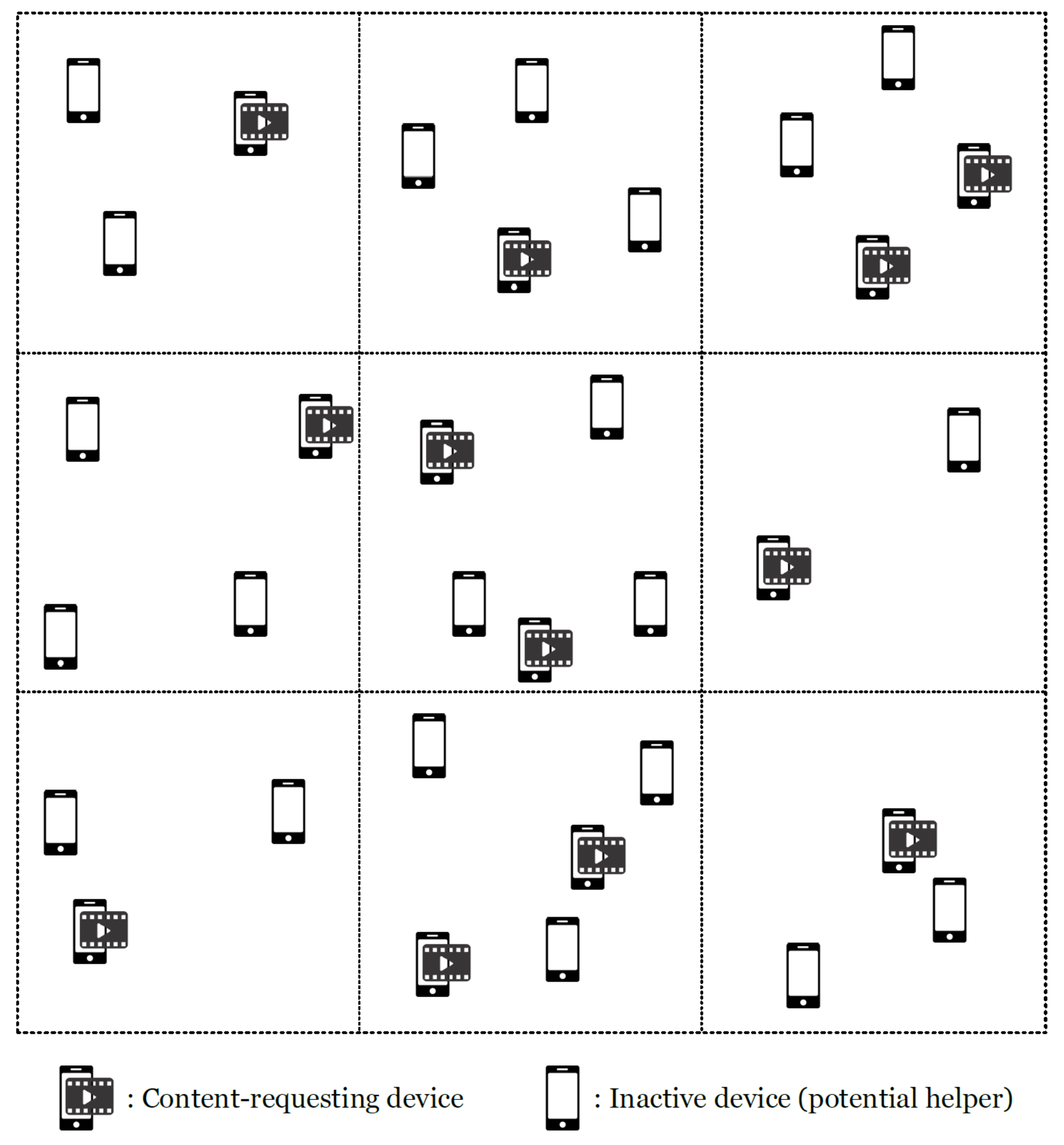}
		\caption{D2D-assisted caching network}
		\label{fig:d2d_model}
		\endminipage
		\vspace{-3mm}
	\end{figure}
	
	\subsection{Wireless Caching Network}
	\label{subsec:caching_network}
	
	We consider a wireless caching network consisting of $M$ caching nodes and $N$ users. 
	Denote each caching node and user by $c_m \in \mathcal{C}$ and $u_n \in \mathcal{U}$, respectively, where $m\in\{1,\cdots ,M \}$, $n\in\{1,\cdots, N \}$, and $\mathcal{C}$ and $\mathcal{U}$ represent the caching node set and the user set, respectively.
	The central server or macro BS has already pushed popular contents during off-peak hours to caching nodes, the storage size of which is finite.
	Each user requests one of the contents in a library $\mathcal{F}$ according to the popularity distribution, e.g., a Zipf distribution, and the requested content can be provided from one of the nearby caching nodes.
	The caching network considered in this paper can be a (BS-assisted) helper network or D2D-assisted network, as shown in Figs. \ref{fig:network_model} and \ref{fig:d2d_model}, respectively.
	In the helper network, helper locations are fixed and usually $M<N$.
	The spatial user distribution follows homogeneous Poisson point processes (PPPs) with intensity $\lambda$.
	In the D2D-assisted network, every device can cache contents. 
	User devices are randomly distributed based on independent PPPs, again with intensity $\lambda$. 
	A few devices request the content, and the remaining inactive devices can act as caching nodes; therefore, usually $M>N$.
	The difference between the setups in Figs. \ref{fig:network_model} and \ref{fig:d2d_model} is explained in detail in Section \ref{sec:numerical_results}.
	
	Because this study is focused on the delivery phase only, content placements in caching nodes are assumed to be already completed. 
	A user needs to choose one of the caching nodes that (i) store the requested content and (ii) have channel conditions that are sufficiently strong to allow successful delivery of the content.
	Therefore, node association becomes a link scheduling problem.
	We consider discretized time slots, i.e., $t\in \{1,2\cdots, \}$, and link scheduling and power allocation are updated for every slot.
	The key difference between this and the standard, well-explored, link scheduling problem is that each receiver has {\em multiple} possible transmitters from which it can obtain the content.
	Note that, at a given time, a caching node can be in the idle state; i.e., it can deliver no content and preserve its power.
	Suppose that every caching node has the same power budget, $q_{\text{max}}$, and each $c_m$ can adjust its transmit power satisfying $0 \leq q_m \leq q_{\text{max}}$.
	In general, user demands are generated asynchronously. 
	Consequently, broadcasting the content to multiple users and cooperation among caching nodes are not allowed in this scheme. 
	Therefore, only one-to-one link scheduling is considered; in other words, each user receives the content from one node and each caching node delivers the signal to only one user.
	
	In this paper, the definitions of the signal link and the interference link are as follows.
	\begin{definition}
		\label{def:sig_link}
		The \textit{signal link} describes the connection from $c_m$ to $u_n$, when $c_m$ caches the content requested by $u_n$, $u_n$ is not scheduled to receive the desired content from another caching node $c_k \neq c_m$, and the averaged SNR at $u_n$ is larger than a minimum required signal-to-noise ratio (SNR) $\Gamma$ dB.
		Given the transmission power range, channel statistics, and noise variance, the constraint on a minimum required SNR can be converted into the cooperation distance $d_s$; in other words, if the distance between $c_m$ and $u_n$ is smaller than $d_s$, $u_n$ can obtain a minimum SNR $\Gamma$ dB from $c_m$.
	\end{definition}
	\begin{definition}
		\label{def:intf_link}
		The \textit{interference link} describes the connection between  $c_m$ and $u_n$, when $u_n$ is not receiving the content from $c_m$, although $c_m$ transmits to another user, and the distance between $c_m$ and $u_n$ is smaller than $d_i$. 
		Here, $d_i$ defines the interference coverage of $c_m$.
	\end{definition}
	Definition \ref{def:sig_link} implies that only caching nodes within radius $d_s$ of an active user can potentially deliver the content successfully. 
	Similarly, the interference effect of the caching node on users at distances larger than $d_i$ is ignored in Definition \ref{def:intf_link}.
	In general, the signal power should be stronger than the interference power; therefore, we assume that  $d_i > d_s$.
	Guidelines for information-theoretically optimum rules concerning which interference contributions should be treated as noise are provided, e.g., in \cite{ITlinQ}. 
	
	We term a pair of $c_m$ and $u_n$ as ``neighboring" when $c_m$ can generate the signal link or the interference link with $u_n$.
	In this sense, $\mathcal{U}_m$ and $\mathcal{H}_n$ represent the active user set neighboring to $c_m$ and the caching node set neighboring to $u_n$, respectively.
	In $\mathcal{U}_m$, users who can construct a signal link with $c_m$ compose the set $\mathcal{V}_m$.
	Similarly, $\mathcal{J}_n$ denotes the caching node set, the members of which can generate a signal link with $u_n$.
	In other words, $c_m$ enables the delivery of the content requested by $u_n$ when $u_n \in \mathcal{V}_m$ and $c_m \in \mathcal{J}_n$. 
	The scheduling indicator $x_{mn} \in \{0, 1\}$ notifies whether $c_m$ schedules $u_n$. 
	
	Then, the link rate between $c_m$ and $u_n$ can be written as
	\begin{equation}
	R_{mn}(t) = \mathcal{B} \log_2 \bigg( 1 + \frac{ \sum_{m\in \mathcal{J}_n} |h_{mn}|^2 \cdot x_{mn} q_{m} }{ \sum_{i \in \mathcal{H}_n} |h_{in}|^2 \sum_{\substack{k\in \mathcal{V}_i \\ k\neq n} } x_{ik} q_{i} + \sigma^2 } \bigg), \label{eq:rate1}
	\end{equation}
	where $h_{mn}$ is the channel gain between $c_m$ and $u_n$, $\mathcal{B}$ is the bandwidth, and $\sigma^2$ is the noise variance.
	Assume that every scheduled user accesses the same bandwidth.
	The data rate of $u_n$ becomes $R_n(\mathbf{x}_{\mathcal{H}_n}, \mathbf{q}_{\mathcal{H}_n}, t ) = \sum_{c_m\in \mathcal{H}_n} R_{mn}(t)$, where $\mathbf{x}_{\mathcal{H}_n}$ and $\mathbf{q}_{\mathcal{H}_n}$ are respectively the scheduling indicator and the power allocation vectors of caching nodes neighboring to $u_n$.
	We denote $\mathbf{x}_{\mathcal{H}_n} = [\mathbf{x}_m : c_m \in \mathcal{H}_n]$, where $\mathbf{x}_m = [x_{mi} : u_i \in \mathcal{U}_m]$, and $x_{mi} \in \{0,1\}$ for all $c_m\in \mathcal{C}$ and $u_i \in \mathcal{U}_m$.
	Similarly, $\mathbf{q}_{\mathcal{H}_n} = [q_m : c_m \in \mathcal{H}_n]$.
	This paper considers one-to-one scheduling only, which indicates that each requesting user should receive the desired content from only one caching node, and each caching node should deliver the content to only one user.
	Therefore, given $u_n$, at most one caching node can deliver the content to $u_n$ for all $c_m \in \mathcal{H}_n$.
	In other words, $\sum_{c_m \in \mathcal{H}_n} x_{mn} \leq 1$ and $\sum_{c_m \in \mathcal{H}_n} \mathcal{I}(R_{mn}(t) > 0) \leq 1$, where $\mathcal{I}(.)$ is the indicator function.
	
	Note that we do not make any assumption that would restrict the network structure (e.g., small cells having their own coverage, femtocells, and clusters) and delivery links (e.g., cellular links, D2D links, and links using of orthogonal resources).
	The system model and problem formulation considered in this paper can be applied to any general cache-assisted content delivery network where delay-sensitive demands are randomly generated and active links interfere with each other. 
	Caching nodes can be small cell BSs, caching helpers, or cache-enabled devices.
	Activated delivery links can be cellular links or D2D links, and it is unimportant whether orthogonal or the same frequency resources are used to satisfy content-requesting users.
	
	\subsection{User Queue Model}
	\label{subsec:queue_model}
	
	Suppose that each content is divided into many chunks and $u_n$ consecutively receives chunks from one node in $\mathcal{J}_n$ in each slot.
	Assuming that the caching nodes and users are not moving, $\mathcal{J}_n$ and $\mathcal{V}_m$ do not change during the considered time frame.
	The number of requested chunks is also random at every slot, and we assume that user demands are accumulated in the user queue.
	
	For each $u_n \in \mathcal{N}$, the queue dynamics in each slot $t \in \{0, 1, \cdots \}$ can be represented as $Q_n(t+1) = \max \{ Q_n(t) - \tilde{\mu}_n(t), 0 \} + a_{n}(t)$,
	where $Q_n(t)$, $a_n(t)$, and $\tilde{\mu}_n(t)$ represent the queue backlog, arrival process and departure process of $u_n$ in slot $t$, respectively.
	Since $\tilde{\mu}_n(t)$ chunks depart before $a_n(t)$ chunks arrive, the queue dynamics can be also re-written using the actual departure process $\mu_n(t)$, which includes implicitly the treatment of empty queues; i.e., $Q_n(t+1) = Q_n(t) - \mu_n(t) + a_n(t)$, where $\mu_n(t) = \min\{\tilde{\mu}_n(t), Q_n(t) \}$.
	The interval of each slot is assumed to be the channel coherence time, $\tau_c$; we further assume a block fading channel, the channel gain of which is static during the unit slot.
	
	The queue backlog $Q_n(t)$ represents 
	the number of chunks requested by $u_n$ but not delivered until slot $t$.
	$a_n(t)$ and $\mu_n(t)$ respectively represent the numbers of chunks requested by $u_n$ and delivered by the associated caching node at slot $t$. 
	For simplicity, we assume that $a_n(t)$ is an i.i.d. uniform random variable; i.e., $a_n \sim \mathcal{U}(0, a_{\text{max}})$. 
	Conversely, $\mu_n(t)$ depends on the link scheduling and power allocation for $u_n$. 
	Therefore, the departure process of $u_n$ is given by
	\begin{equation}
	\mu_{n}(t) \triangleq \mu_n(\mathbf{x}_{\mathcal{H}_n}, \mathbf{q}_{\mathcal{H}_n}, t) = \min \bigg\{ \bigg\lfloor \frac{\tau_c R_{n}(\mathbf{x}_{\mathcal{H}_n}, \mathbf{q}_{\mathcal{H}_n}, t) }{S} \bigg\rfloor, Q_n(t) \bigg\}, \label{eq:departure}
	\end{equation}
	where $S$ is the size of a chunk.
	Some chunks can be only partially delivered, as the channel condition varies and link scheduling is updated at every time slot $t$. 
	However, since partial chunk transmission is meaningless in our algorithm, a flooring operation is applied in \eqref{eq:departure}.
	
	\section{Joint link scheduling and power allocation for content delivery in wireless caching networks}
	\label{sec:joint_scheduling}
	
	This paper considers two performance metrics in this model: 1) average service delay, i.e., the waiting time of user demands to be served, and 2) power efficiency.
	For achieving both goals, decision parameters of $x_{mn}$ and $q_m$ for all $c_m \in \mathcal{C}$ and $u_n \in \mathcal{U}$ should be carefully determined. 
	The optimization problem that maximizes the long-term time-averaged power efficiency, while limiting the time-averaged service delay, can be formulated as
	\begin{align}
	\{\mathbf{x}, \mathbf{q} \} &= \underset{x_{mn}, q_m, \forall c_m, \forall u_n}{\arg\min}~ \underset{T \rightarrow \infty }{\lim} ~\frac{1}{T} \sum_{t=1}^T \mathbb{E} \bigg[ \sum_{m = 1}^M q_m \sum_{n \in \mathcal{U}_m} x_{mn} \bigg] \label{op1:power_eff} \\
	&~~~\text{s.t.}~~\underset{T\rightarrow \infty}{\lim}~\frac{1}{T} \sum_{t=1}^T \mathbb{E} \bigg[ \sum_{n=1}^N Q_n(t) \bigg] < \infty \label{op1:const:queue} \\
	&~~~~~~~~~ 0\leq q_m\leq q_{\text{max}},~\forall c_m \in \mathcal{C} \label{op1:const:power} \\
	&~~~~~~~~~ x_{mn} \in \{0,1 \},~\forall c_m \in \mathcal{C},~\forall u_n \in \mathcal{U} \label{op1:const:link} \\
	&~~~~~~~~~ \sum_{m=1}^M x_{mn} \leq 1,~\forall u_n \in \mathcal{U} \label{op1:const:one_user} \\
	&~~~~~~~~~ \sum_{n=1}^N x_{mn} \leq 1,~\forall c_m \in \mathcal{C}, \label{op1:const:one_helper}
	\end{align}
	where $\mathbf{x} = [x_{mn} : c_m \in \mathcal{C}, u_n \in \mathcal{U}_m]$, and $\mathbf{q} = [q_m : c_m \in \mathcal{C}]$.
	
	Specifically, the expectations of both \eqref{op1:power_eff} and \eqref{op1:const:queue} are with respect to random channel realizations. 
	Constraint \eqref{op1:const:queue} pursues the strong stability of the user queueing system, and constraint \eqref{op1:const:power} limits the peak power, which could reflect either constraints imposed by a frequency regulator, such as the Federal Communications Commission (FCC), or the limitations of the hardware, such as the power amplifier.
	The scheduling indicator is zero or one in \eqref{op1:const:link}, and the one-to-one link scheduling is guaranteed by \eqref{op1:const:one_user} and \eqref{op1:const:one_helper}.
	
	As mentioned previously, the waiting time in the user queue of demands for service is closely related to the service delay. 
	According to Little's theorem \cite{LittlesThm}, the averaged queueing delay is proportional to the average queue length; therefore, the long-term time-averaged service delay can be limited by taking constraint \eqref{op1:const:queue}.
	Based on Lyapunov optimization theory, the upper bound on the time-averaged queue length is also derived by using an algorithm that minimizes the Lyapunov drift \cite{Lyapunov}. 
	Finally, the solution of the problem represented by \eqref{op1:power_eff}--\eqref{op1:const:one_helper} averts the excessive accumulation of user demands in the queue by achieving queue stability in \eqref{op1:const:queue}.
	
	For pursuing the stability of queueing systems of all users, the optimization problem represented by \eqref{op1:power_eff}--\eqref{op1:const:one_helper} can be solved based on Lyapunov optimization theory. 
	Let $\mathbf{Q}(t)$ denote the column vector of $Q_n(t)$ of all users at time $t$, and define the quadratic Lyapunov function $L(\mathbf{Q}(t))$ as 
	\begin{equation}
	L(\mathbf{Q}(t)) = \frac{1}{2} \sum_{n=1}^N (Q_n(t))^2.
	\end{equation}
	Then, let $\Delta(.)$ be a conditional quadratic Lyapunov function that can be formulated as $\mathbb{E}[L(\mathbf{Q}(t+1)) - L(\mathbf{Q}(t)) | \mathbf{Q}(t)]$, i.e., the drift on $t$. 
	The dynamic policy is designed to achieve queue stability in \eqref{op1:const:queue} by observing the current queue state $\mathbf{Q}(t)$ and determining link scheduling and power allocation to minimize an upper bound on \textit{drift-plus-penalty} \cite{Lyapunov}:
	\begin{equation}
	\Delta (\mathbf{Q}(t)) - V \mathbb{E} \bigg[ \sum_{m=1}^M q_m \sum_{u_n \in \mathcal{U}_m} x_{mn} \Big| \mathbf{Q}(t) \bigg], \label{eq:dpp}
	\end{equation}
	where $V$ is an importance weight for power efficiency.
	
	First, the upper bound on the drift can be found in the Lyapunov function.
	\begin{align}
	L(\mathbf{Q}(t+1)) - L(\mathbf{Q}(t)) &= \frac{1}{2} \sum_{n=1}^{N} \bigg[ Q_n(t+1)^2 - Q_n(t)^2 \bigg] \\
	&\leq \frac{1}{2} \sum_{n=1}^N \Big[ a_n(t)^2 + \mu_n(t)^2 \Big] + \sum_{n=1}^N Q_n(t) (a_n(t) - \mu_n(t)). 
	\end{align}
	Then, the upper bound on the conditional Lyapunov drift is obtained as
	\begin{align}
	\Delta(\mathbf{Q}(t)) = \mathbb{E}[ L(\mathbf{Q}(t+1)) - L(\mathbf{Q}(t)) | \mathbf{Q}(t) \leq C + \mathbb{E} \bigg[ \sum_{n=1}^N Q_n(t) (a_n(t) - \mu_n(t)) \Big| \mathbf{Q}(t) \bigg],
	\end{align}
	where 
	\begin{equation}
	\frac{1}{2} \mathbb{E}\bigg[ \sum_{n=1}^N \Big[ a_n(t)^2 + \mu_n(t)^2 \Big] \Big| \mathbf{Q}(t) \bigg] \leq C,
	\end{equation}
	which assumes that the arrival and departure process rates are upper bounded.
	According to \eqref{eq:dpp}, minimizing the bound on drift is consistent with maximizing 
	\begin{equation}
	\mathbb{E} \bigg[ \sum_{n=1}^N Q_n(t) \mu_n(t) \Big| \mathbf{Q}(t) \bigg] - V \mathbb{E} \bigg[ \sum_{m=1}^M q_m \sum_{u_n \in \mathcal{U}_m} x_{mn} \Big| \mathbf{Q}(t) \bigg], \label{eq:max-weight}
	\end{equation}
	because $C$ is a constant and $a_n(t)$ for all $u_n\in\mathcal{U}$ is not controllable.
	We now use the concept of opportunistically minimizing the expectations; therefore, \eqref{eq:max-weight} is minimized by an algorithm that observes the current queue state $\mathbf{Q}(t)$ and determines $\mathbf{x}$ and $\mathbf{q}$ to maximize 
	\begin{equation}
	\sum_{n=1}^N Q_n(t) \mu_n(t) - V \sum_{m=1}^M q_m \sum_{u_n \in \mathcal{U}_m} x_{mn}. \label{eq:dpp_upper}
	\end{equation}
	
	System parameter $V$ in \eqref{eq:dpp_upper} is a weight factor for the term representing power efficiency. 
	$V$ can be regarded as the parameter for controlling the tradeoff between power efficiency and service delay, which captures the fact that a helper can clear many backlogs or deliver a relatively small number of requested chunks using little transmit power under the given channel condition. 
	The appropriate initial value of $V$ needs to be obtained empirically, because it depends on the distributions of caching nodes and users, channel environments, and arrival rate.
	In addition, $V\geq 0$ should be satisfied.
	If $V<0$, the optimization goal is converted into maximizing the summation of the power consumption.
	Moreover, in the case where $V=0$, the user aims only at clearing queue backlogs without considering power efficiency.
	However, when $V\rightarrow \infty$, users do not consider the queue state and pursue minimization of the power consumption only.
	
	According to \eqref{eq:dpp_upper}, the utility function representing the negative sign of the upper bound on drift-plus-penalty can be written as
	\begin{align}
	F(\mathbf{x}, \mathbf{q}, t) &= \sum_{n=1}^{N} Q_n(t) \mu_n ( \mathbf{x}_{\mathcal{H}_n}, \mathbf{q}_{\mathcal{H}_n}, t ) - V \cdot \sum_{m=1}^M q_{m} \sum_{u_n \in \mathcal{U}_m} x_{mn} \\
	&= \sum_{n=1}^{N} Q_n(t) \mu_n ( \mathbf{x}_{\mathcal{H}_n}, \mathbf{q}_{\mathcal{H}_n}, t ) - V \cdot \sum_{n=1}^N \sum_{c_m \in \mathcal{H}_n} q_{m} x_{mn} \\
	&= \sum_{n=1}^N \tilde{f}_n ( \mathbf{x}_{\mathcal{H}_n}, \mathbf{q}_{\mathcal{H}_n}, t ).
	\end{align}
	For simplicity, notations for the dependency of all parameters on $t$ are omitted in the remaining sections because link scheduling and power allocation are determined in every different slot. 
	
	Therefore, according to Lyapunov optimization theory \cite{Lyapunov}, the problem of \eqref{op1:power_eff}--\eqref{op1:const:one_helper} can be converted into the \textit{min-drift-plus-penalty} problem:
	\begin{align}
	\{ \mathbf{x}^{\star}, \mathbf{q}^{\star} \} =~ &\underset{ \mathbf{x}, \mathbf{q} }{\arg\max}~ \sum_{n=1}^N \tilde{f}_n ( \mathbf{x}_{\mathcal{H}_n}, \mathbf{q}_{\mathcal{H}_n} ) \label{op2:opt} \\
	\text{s.t.}~~&\eqref{op1:const:power}-\eqref{op1:const:one_helper}. \label{op2:constraints}
	\end{align}
	Here, $\mathbf{x}^{\star}$ and $\mathbf{q}^{\star}$ are the optimal link scheduling and power allocation vectors, respectively.
	Since a certain caching node $c_m$ interferes with users receiving contents from other active caching nodes, i.e., $c_i \neq c_m$, the objective function of \eqref{op2:opt} is not separable.
	Therefore, distributed decisions on $\mathbf{x}^{\star}$ and $\mathbf{q}^{\star}$ are difficult in this formulation. 
	
	Solutions of the problem of \eqref{op2:opt}--\eqref{op2:constraints} can be obtained by relaxing the constraints of \eqref{op1:const:one_user} and \eqref{op1:const:one_helper}.
	First, we consider discrete power levels, i.e., $q_{m} \in \{0, P_1, \cdots, P_L \}$, and  
	$P_l > 0$ for all $l\in\{1,\cdots,L \}$. 
	When $c_m$ is in the idle state, $\sum_{u_n \in \mathcal{U}_m} x_{mn} = 0$, $q_m = 0$.
	This is reasonable, because practical power control uses discrete levels. 
	Second, to design a distributed delivery scheme, we suppose that each $c_m$ determines $\mathbf{x}_m$ and $\mathbf{q}_m$ in a probabilistic manner; therefore, the goal of the content delivery scheme becomes to find probability distributions of link scheduling and power allocation at all caching nodes, i.e., $p(\mathbf{x}_m, q_m)$ for all $c_m \in \mathcal{C}$.
	Assuming that $c_m$ is not in the idle state, i.e., $\sum_{u_n \in \mathcal{U}_m} x_{mn} = 1$, the most probable decisions on link scheduling and power allocation are determined by 
	\begin{equation}
	\{ \mathbf{x}_m^*, q_m^* \} = \underset{u_n\in \mathcal{U}_m, l\in\{1,\cdots,L \} }{\arg\max}~\mathcal{P}_{mnl}, \label{eq:prob_dec}
	\end{equation}
	where $\mathcal{P}_{mnl} = \mathrm{Pr} \{ x_{mn}=1, \sum_{u_i \in \mathcal{U}_m} x_{mi} = 0, q_m = P_l \}$,
	and the optimal value of \eqref{eq:prob_dec} is $p_m^*$.
	Then, compare with the case where $c_m$ is in the idle state to make the final decisions:
	\begin{equation}
	\{ \mathbf{x}_m^{\star}, q_m^{\star} \} = 
	\begin{cases}
	\{ \mathbf{x}_m^*, q_m^* \},~~& p_m^* > \mathrm{Pr}\Big\{ \sum_{n \in \mathcal{U}_m} x_{mn} = 0 \Big\} \\
	\{\mathbf{0}_{|\mathcal{U}_m| \times 1}, 0 \},~~& \text{otherwise}
	\end{cases}, \label{eq:decision_process}
	\end{equation}
	where $\{ \mathbf{x}_m^*, q_m^* \} = \{\mathbf{0}_{|\mathcal{U}_m| \times 1}, 0 \}$ signifies that $c_m$ is in the idle state, where $\mathbf{0}_{|\mathcal{U}_m|\times 1}$ is the zero vector with size $|\mathcal{U}_m| \times 1$.
	Thus, there are $(L\cdot |\mathcal{V}_m| + 1)$ possible decisions. 
	
	According to the above decision process, every caching node can either schedule only one user or be in the idle state; therefore, constraint \eqref{op1:const:one_helper} can be removed.
	In addition, constraint \eqref{op1:const:one_user} can be combined with the objective function \eqref{op2:opt} by using the indicator function. 
	Finally, the problem of \eqref{op2:opt}--\eqref{op2:constraints} can be re-written as 
	\begin{align}
	&p(\mathbf{x}, \mathbf{q} ) = \underset{ p(\mathbf{x}_m, q_m) }{\arg\max}~ \sum_{n=1}^N  f_n( \mathbf{x}_{\mathcal{H}_n}, \mathbf{q}_{\mathcal{H}_n} ) \label{op3:opt} \\
	&~~~~~~~\text{s.t.}~~q_m \in \{P_1,\cdots, P_L \},~\forall c_m \in \mathcal{C} \label{op3:q_m} \\
	&~~~~~~~~~~~~~x_{mn} \in \{0,1 \},~\forall c_m \in \mathcal{C}~ \forall u_n \in \mathcal{U} \label{op3:x_mn} \\
	&~~~~~~~~~~~~~\eqref{eq:prob_dec}, \eqref{eq:decision_process},
	\end{align}
	where $f_n( \mathbf{x}_{\mathcal{H}_n}, \mathbf{q}_{\mathcal{H}_n} ) = \tilde{f}_n ( \mathbf{x}_{\mathcal{H}_n}, \mathbf{q}_{\mathcal{H}_n} ) \cdot \mathcal{I}\Big( \sum_{c_m\in \mathcal{H}_n} x_{mn} \leq 1 \Big)$.
	The goal of the above problem becomes to find the marginal probabilities $p(\mathbf{x}_m, q_m)$ for all caching nodes $c_m\in \mathcal{C}$, and this goal can be achieved by using the BP algorithm.
	
	\begin{figure} [t!]
		\centering
		\includegraphics[width=0.35\textwidth]{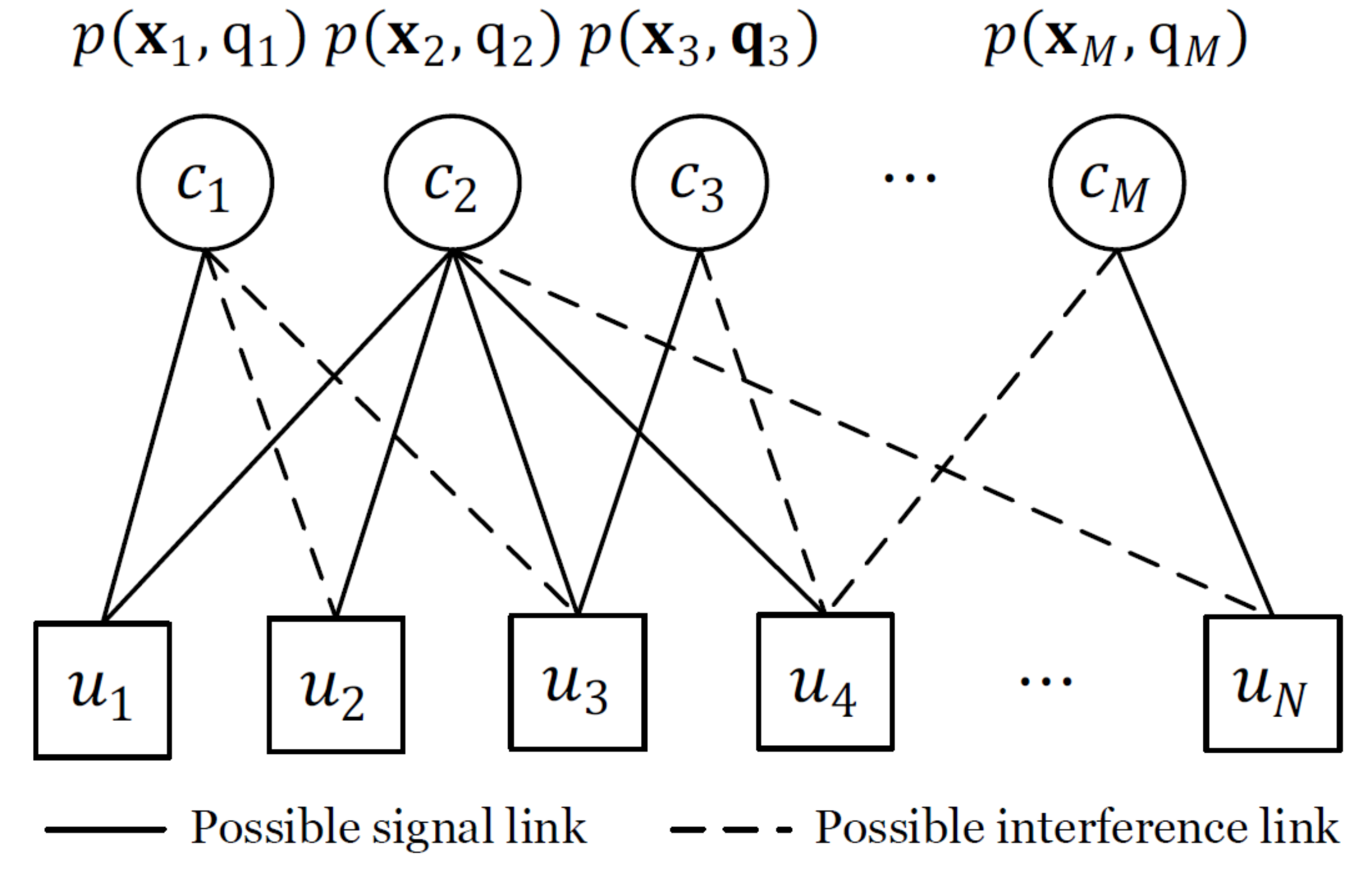}
		\caption{Factor graph consisting of caching helpers and users}
		\label{fig:factor_graph}
		\vspace{-3mm}
	\end{figure}
	
	\section{Belief Propagation Algorithm}
	\label{sec:BP}
	
	This section explains how the optimization problem represented by \eqref{op3:opt}--\eqref{op3:x_mn} can be solved by using the BP algorithm.
	We define the probability distributions of all possible $\mathbf{x}_m$ and $q_{m}$ for all $c_m \in \mathcal{C}$ and $u_n \in \mathcal{U}$ with a constant $\delta > 0$ as 
	\begin{align}
	p(\mathbf{x}, \mathbf{q}) = \frac{1}{Z} \exp \Big(\delta F(\mathbf{x}, \mathbf{q}) \Big) = \frac{1}{Z} \prod_{n=1}^N \exp\Big( \delta f_n(\mathbf{x}_{\mathcal{H}_n},\mathbf{q}_{\mathcal{H}_n}) \Big),
	\end{align}
	where $Z$ is a normalization factor called the partition function of $\delta$. 
	The goal is to find $p(\mathbf{x}_m, q_m)$ for all $c_m \in \mathcal{C}$ to decide link scheduling and power allocation at every caching node in a distributed manner.
	The marginal distribution of $p(\mathbf{x}, \mathbf{q})$ with respect to the decision variables $\mathbf{x}_m$ and $q_m$, i.e., $p(\mathbf{x}_m, q_m)$, can be estimated by the BP algorithm.
	According to a standard result of large deviations \cite{large_deviation}, the optimal decisions to maximize the utility function $F(\mathbf{x}, \mathbf{q})$ as $\delta \rightarrow \infty$ are 
	\begin{equation}
	\underset{\delta \rightarrow \infty}{\lim}\{ \hat{\mathbf{x}}, \hat{\mathbf{q}} \} = \underset{\mathbf{x}, \mathbf{q}}{\arg\max}~ F(\mathbf{x}, \mathbf{q}).
	\end{equation}
	Therefore, we can estimate the marginal expectations of the probability distribution $p(\mathbf{x}, \mathbf{q})$ for large $\delta$, and $c_m$ can decide link scheduling and power allocation based on the marginalized $p(\mathbf{x}_m, q_m)$.
	
	A bipartite graph $G=(V,E)$, called the factor graph, is constructed to represent the network topology, where the vertex set $V$ consists of $M$ caching nodes and $N$ users, as shown in Fig. \ref{fig:factor_graph}. 
	In the factor graph, caching nodes are variable nodes and users are factor nodes.
	There exists an edge $(c_m, u_n) \in E$ if $c_m$ and $u_n$ neighbor each other, i.e., $c_m$ can interfere with $u_n$. 
	In other words, an edge $(c_m, u_n) \in E$ implies $c_m \in \mathcal{H}_n$ and $u_n \in \mathcal{U}_m$.
	In the factor graph, only caching nodes that store at least one or more contents requested by neighboring users are considered, and only users who can find at least one or more neighboring caching nodes storing the requested content are considered.
	
	In the BP algorithm, variable nodes and factor nodes iteratively exchange the belief messages along the edges of the factor graph.
	The belief messages of the variable node representing $c_m$ transmitted to and received from neighboring factor nodes deliver estimates of $p(\mathbf{x}_m, q_m)$.
	Denote the belief message delivered from $u_n$ to $c_m$ at iteration $i$ by $p^i_{n \rightarrow m}(\mathbf{x}_m, q_m)$ and the reverse message that $c_m$ passes to $u_n$ by $p^i_{n \leftarrow m}(\mathbf{x}_m, q_m)$.
	After the messages are exchanged between users and caching nodes for a fixed number of iterations, the final decision is made at each node $c_m$ to compute the marginalized distribution $p(\mathbf{x}_m, q_m)$.
	The BP algorithm steps are as follows.
	
	\begin{enumerate}
		\item Initialization: 
		Before the iterations are started, the initial values are given to $p^1_{n\leftarrow m}(\mathbf{x}_m, q_m)$ for all $c_m \in \mathcal{C}$ and $u_n \in \mathcal{U}$.
		The variable node representing $c_m$ has $(L\cdot |\mathcal{V}_m| + 1)$ decision parameters, including $L\cdot |\mathcal{V}_m|$ active links with different scheduling and power allocations, and one idle state.
		We suppose that every possible decision on link scheduling and power allocation at each caching node follows a uniform distribution at $i=1$; therefore, $p^1_{n\leftarrow m}(x_{mn}=1, \sum_{ \substack{ u_k \in \mathcal{V}_m \\ k \neq n } } x_{mk} = 0, q_{m} = P_l ) = \frac{1}{L \cdot |\mathcal{V}_m|+1}$, for all $c_m \in \mathcal{C}$, $u_n \in \mathcal{U}_m$, and $l \in \{1,\cdots,L \}$.
		Similarly, the initial probability of being in the idle state is $p_{n \leftarrow m}^1 (\sum_{u_k \in \mathcal{V}_m} x_{mk} = 0) = \frac{1}{L \cdot |\mathcal{V}_m|+1}$. 
		
		\item Factor node update: 
		$u_n$ (i.e., factor node $n$) updates the belief message $p^i_{n \rightarrow m}(\mathbf{x}_m, q_m)$ and sends it to $c_m$ (i.e., variable node $m$), where $c_m \in \mathcal{H}_n$. 
		$u_n$ computes $p^i_{n \rightarrow m}(\mathbf{x}_m, q_m)$ based on the messages received from all caching nodes $c_k \in \mathcal{H}_n \setminus \{c_m \}$, i.e., $p^{i}_{n \leftarrow k}(\mathbf{x}_k, q_k)$, as given by
		\begin{equation}
		p^i_{n \rightarrow m}(\mathbf{x}_m, q_m) = \mathbb{E} \Big[ \exp \Big( \delta f_n( \mathbf{x}_{\mathcal{H}_n}, \mathbf{q}_{\mathcal{H}_n} ) \Big) \Big| \mathbf{x}_m, q_m \Big]. \label{eq:factor_update}
		\end{equation}
		The expectation represented by \eqref{eq:factor_update} is with respect to $p^{i}_{n \leftarrow k}(\mathbf{x}_k, q_k)$ for all $c_k \in \mathcal{H}_n \setminus \{c_m \}$.
		When computing \eqref{eq:factor_update}, three different cases need to be considered: 1) $c_m$ and $u_n$ generate the signal link; i.e., $c_m$ delivers the content to $u_n$; 2) $c_m$ and $u_n$ generate the interference link, and $u_n$ receives the content from another neighboring caching node $c_j \neq c_m$; and 3) $u_n$ cannot receive any content from neighboring caching nodes. 
		Note that the third case includes both the situation where $\sum_{c_s \in \mathcal{H}_n} x_{sn} = 0$ and that where $\mathcal{I}\Big( \sum_{c_m\in \mathcal{H}_n} x_{mn} \leq 1 \Big) = 0$, representing the case when no neighboring caching node schedules $u_n$ and when two or more caching nodes schedule $u_n$ at the same time, respectively.
		
		$p^i_{n\rightarrow m}( x_{mn}=1, \sum_{u_k \in \mathcal{U}_m, k \neq n} x_{mk} = 0, q_m = P_l )$ represents the first case, where $c_m$ schedules $u_n$ with the transmit power $P_l$ and is updated by averaging the data rates given by
		\begin{equation}
		\mathcal{B} \log_2 \bigg( 1 + \frac{ |h_{mn}|^2 \cdot q_{m} }{ \sum_{\substack{c_s \in \mathcal{H}_n \\ s \neq m}} |h_{sn}|^2 q_s \sum_{\substack{u_v \in \mathcal{V}_s \\ v\neq n} } x_{sv} + \sigma^2 } \bigg),
		\end{equation}
		with respect to $\mathbf{x}_s$ and $q_s$ for all $c_s \in \mathcal{H}_n \setminus \{c_m \}$.
		
		Meanwhile, $p^i_{n \rightarrow m}( x_{mk}=1, \sum_{u_w \in \mathcal{U}_m, w \neq k} x_{mw} = 0, q_m = P_l )$ for $u_k\in \mathcal{V}_m \setminus \{u_n\}$ is the probability that $c_m$ schedules $u_k$ with transmit power $P_l$ and interferes with $u_n$.
		This probability represents the second case, where the signal link between $u_n$ and another neighboring caching node $c_j$ is generated and can be obtained by averaging the data rates given by
		\begin{equation}
		\mathcal{B} \log_2 \Bigg( 1 + \frac{ \sum_{\substack{c_j\in \mathcal{J}_n \\ j\neq m }} |h_{jn}|^2 q_{j} \cdot \mathcal{I} \Big( \sum_{\substack{c_j\in \mathcal{J}_n \\ j\neq m }} x_{jn} \leq 1 \Big)  }{ |h_{mn}|^2 q_{k} + \sum\limits_{\substack{c_s \in \mathcal{H}_n \\ s \neq m,j }} |h_{sn}|^2 q_s \sum\limits_{\substack{u_v \in \mathcal{V}_s \\ v\neq n} } x_{sv} + \sigma^2 } \Bigg),
		\end{equation}
		with respect to $\{ \mathbf{x}_j, q_j \}$ and $\{ \mathbf{x}_s, q_s \}$ for all $c_j \in \mathcal{J}_n$, $j\neq m$, and $c_s \in \mathcal{H}_n$, $s \neq m, j$.
		
		Finally, in the third case, in which $u_n$ cannot receive any content from neighboring caching nodes, zero throughput is achieved at $u_n$.
		
		\item Variable node update: 
		In every iteration, each caching node sends updated messages to its neighboring users after receiving belief messages from neighboring users.
		
		$c_m$ updates the belief message $p^{i+1}_{n \leftarrow m}(\mathbf{x}_m, q_m)$ by using received messages $p^i_{j \rightarrow m}(\mathbf{x}_m, q_m)$ for all $u_j \in \mathcal{U}_m,~j \neq n$ and sends it to factor node $u_n\in \mathcal{U}_m$:
		\begin{equation}
		p^{i+1}_{n \leftarrow m}(\mathbf{x}_m, q_m) = \frac{1}{Z} \prod_{\substack{u_j\in \mathcal{U}_m \\ j\neq n}} p^i_{j \rightarrow m}(\mathbf{x}_m, q_m).
		\end{equation}
		The updates of belief messages at every factor node and variable node are iteratively performed.
		
		\item Final solution: 
		After the predetermined $I$ iterations, the final decisions at every caching node $c_m \in \mathcal{C}$ can be made based on received messages from neighboring users, as given by
		\begin{equation}
		p^I_{m}(\mathbf{x}_m, q_m) = \frac{1}{Z} \prod_{u_j\in \mathcal{U}_m } p^I_{j \rightarrow m}(\mathbf{x}_m, q_m).
		\end{equation}
	\end{enumerate}
	
	Based on $p^I_{m}(\mathbf{x}_m, q_m)$ for all $c_m \in \mathcal{C}$, each caching node can schedule one of the neighboring users and determine an appropriate power level that provides the largest probability.
	
	\section{Matching Algorithm for One-to-One Link Scheduling}
	\label{sec:matching}
	
\begin{figure}
		\centering
		\begin{subfigure}[b]{0.25\textwidth}
			\centering
			\includegraphics[width=\textwidth]{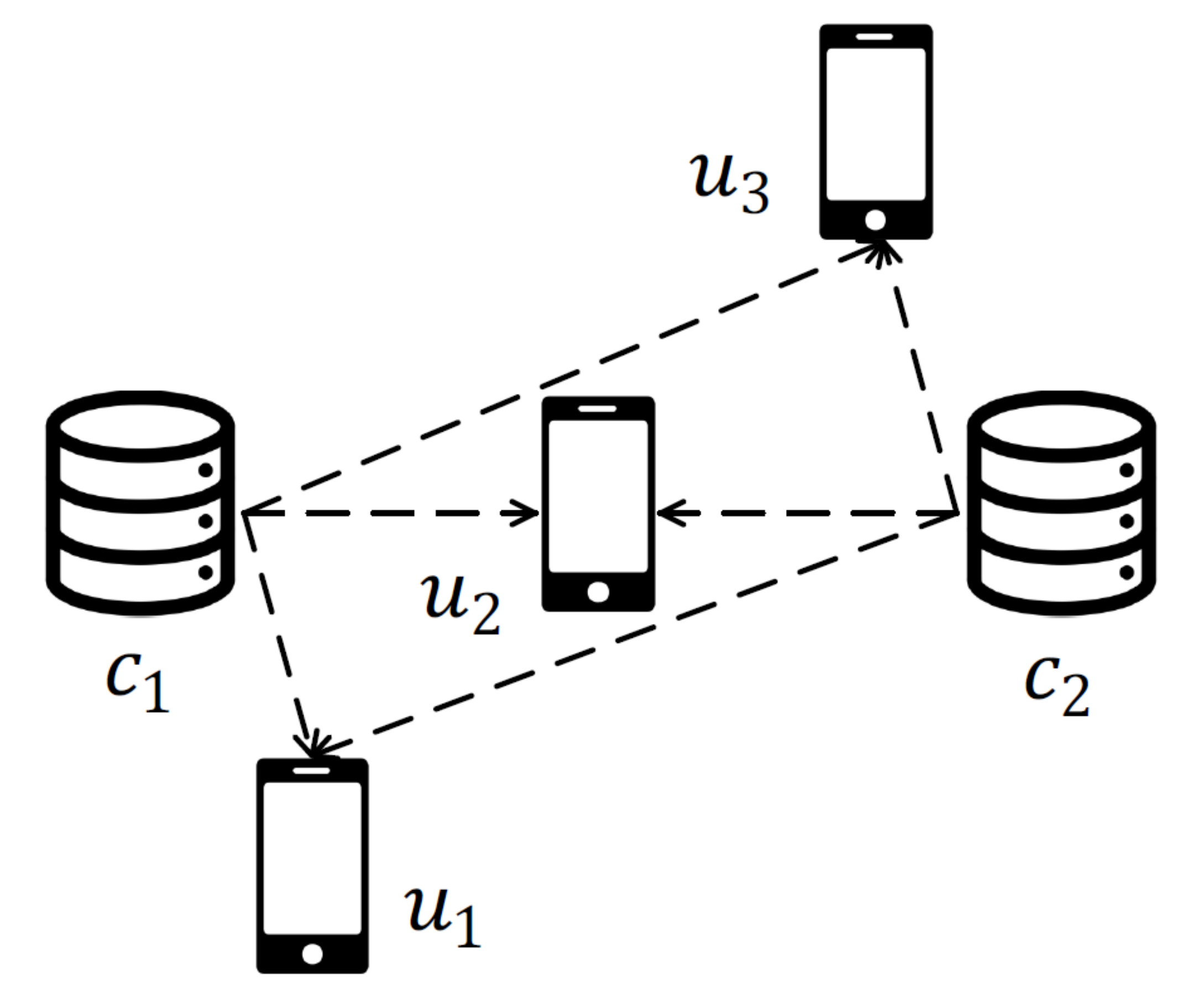}
			\caption{Example of network}
			\label{subfig:ex_topology}
		\end{subfigure}
		\hfill
		\begin{subfigure}[b]{0.25\textwidth}
			\centering
			\includegraphics[width=\textwidth]{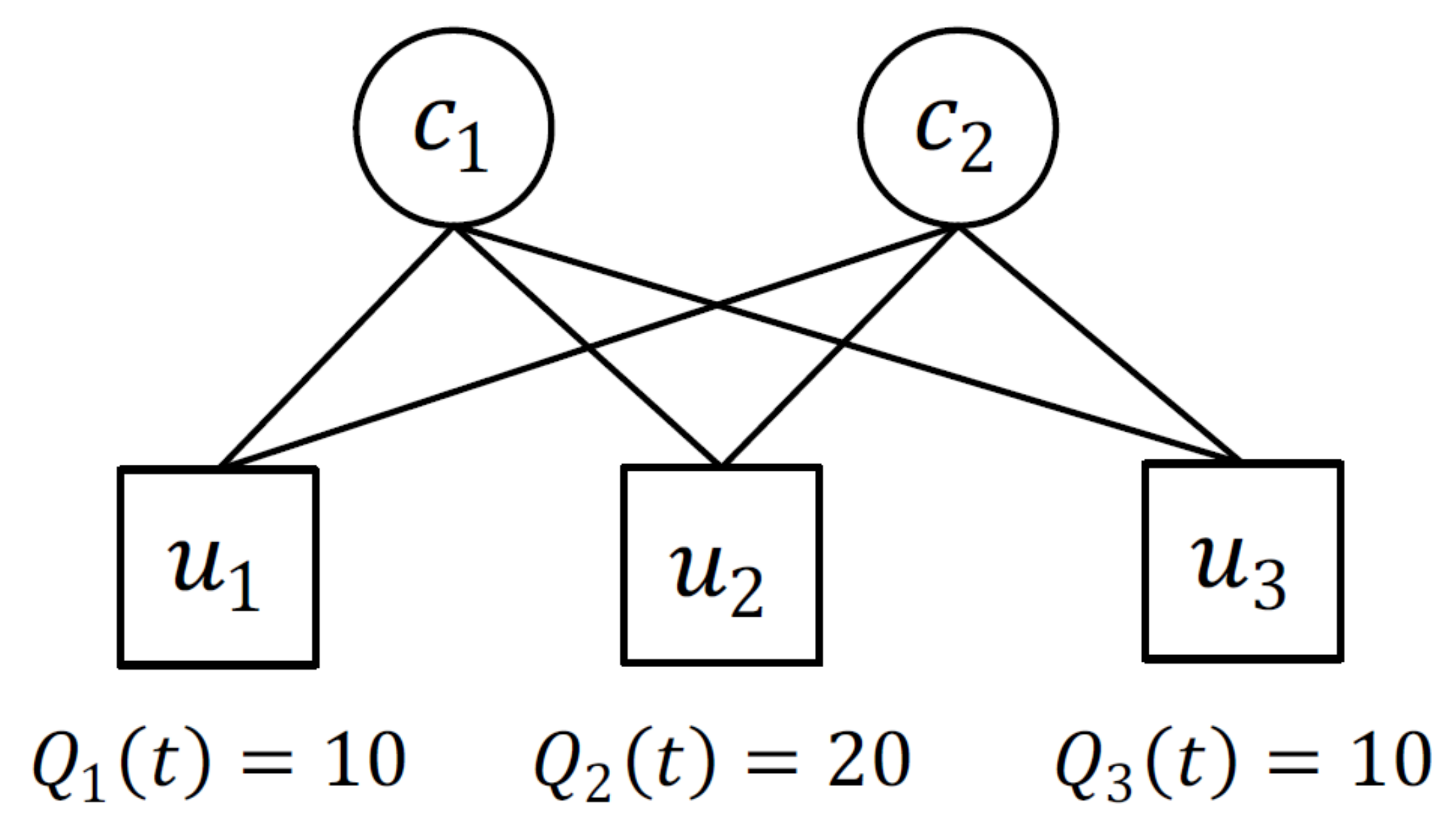}
			\caption{Example of factor graph}
			\label{subfig:ex_graph}
		\end{subfigure}
		\caption{Example of the factor graph for the belief propagation algorithm}
		\label{fig:example}
		\vspace{-3mm}
	\end{figure}

	The BP algorithm proposed in Section \ref{sec:BP} is well suited to estimate the marginal probability distributions of $p(\mathbf{x}_m, q_m)$ for all $c_m \in \mathcal{C}$ and to determine link scheduling and power allocation jointly; however, there are two critical problems.
	According to \cite{BP_cycles}, it is well-known that the BP algorithm is suboptimal when cycles exist in the factor graph; i.e., a belief message transmitted from any node can return to that node within a few iterations.
	In addition, when the length of the cycle (i.e., the number of edges of which the cycle consists) is small, belief messages for nodes in the cycle are not improved over iterations; they could even be degraded. 
	For example, the network topology in Fig. \ref{subfig:ex_topology} can be represented by Fig. \ref{subfig:ex_graph}, and there are two cycles in the factor graph constructed by caching nodes and users.
	The first cycle consists of $c_1$, $u_1$, $c_2$, and $u_2$ and the second cycle of $c_1$, $u_2$, $c_2$, and $u_3$.
	Similarly to this example, the factor graph representing the interference network usually has many cycles with short lengths. 
	
	In addition, because the BP algorithm is a distributed decision method, the updates of belief messages at each factor node or variable node are performed without exact knowledge of other nodes.
	As explained in Section \ref{sec:BP}, the belief messages $p^i_{n \rightarrow m}( \mathbf{x}_m, q_m)$ are updated by averaging messages received from other variable nodes $c_k \in \mathcal{J}_n \setminus \{c_m\}$; therefore, factor node $n$ computes the message to be delivered to variable node $m$ by using imperfect probability distributions for link scheduling and power allocation at other neighboring caching nodes.
	Especially during the first iteration, the belief messages from factor nodes to variable nodes could be incorrectly biased, because the initial messages $p^1_{n\leftarrow m }$ for all $c_m \in \mathcal{C}$ and $u_n \in \mathcal{U}_m$ are uniformly initialized.
	If a very large amount of power is required or a little power is sufficient to deliver the content at the caching node side, this uniform initialization can cause users neighboring to this caching node to underestimate or overestimate interference effects from the caching node.
	Further, although we allow every user to be scheduled by only one caching node by addressing \eqref{op3:opt}, the BP algorithm frequently provides the result that multiple caching nodes are willing to deliver the content to the identical user at the same time. 
	In the next subsection, an example of this problem is introduced.
	
	\subsection{Problem of Belief Propagation Algorithm}
	\label{subsec:prob_BP}
	
	Consider an example network consisting of three users and two caching nodes, as illustrated in Fig. \ref{subfig:ex_topology}, the corresponding factor graph of which is shown in Fig. \ref{subfig:ex_graph}. 
	Assume that the queue lengths of all the users are given by $Q_1(t)=10$, $Q_2(t)=20$, and $Q_3(t)=10$. 
	$u_1$ is closer to $c_1$ than to $c_2$, and therefore, the channel strength between $c_1$ and $u_1$ is larger than that between $c_2$ and $u_1$.
	Similarly, the channel gain from $c_2$ to $u_3$ is stronger than that from $c_1$ to $u_3$.
	Meanwhile, the channel strengths from $c_1$ to $u_2$ and from $c_2$ to $u_2$ are almost the same.
	Since $Q_2(t)$ is larger than $Q_1(t)$ and $Q_3(t)$, $u_2$ sends the relatively strong messages $p^i_{2\rightarrow 1}(\mathbf{x}_1, q_1)$ and $p^i_{2\rightarrow 2}( \mathbf{x}_2, q_2)$ in which $p(x_{12})$ and $p(x_{22})$ are very large, i.e., scheduling of $u_2$ is recommended for both $c_1$ and $c_2$, as compared to messages from $u_1$ and $u_3$ to $c_1$ and $c_2$.
	$u_1$ and $u_3$ are near to $c_1$ and $c_2$, respectively, and $Q_1(t)$ and $Q_2(t)$ are not small; therefore, $p^i_{1\rightarrow 1}(\mathbf{x}_1, q_1)$ and $p^i_{3\rightarrow 2}(\mathbf{x}_2, q_2)$ also assert scheduling of $u_1$ at $c_1$ and of $u_3$ at $c_2$.
	Conversely, $u_3$ is far from $c_1$, and therefore, $p^i_{3\rightarrow 1}(\mathbf{x}_1, q_1)$ notifies that $c_1$ should preferable schedule $u_1$ or $u_2$.
	However, the computation of $p^{i+1}_{2\leftarrow 1}(\mathbf{x}_1, q_1)$ at $c_1$ is based only on $p^i_{1\rightarrow 1}(\mathbf{x}_1, q_1)$ and $p^i_{3\rightarrow 1}(\mathbf{x}_1, q_1)$; therefore,  $p^{i+1}_{2\leftarrow 1}(\mathbf{x}_1, q_1)$ delivers the information that the probability that $c_1$ will schedule $u_1$ is high.
	Similarly, the probability that $u_3$ will be scheduled for $c_2$ is very large in $p^{i+1}_{2\leftarrow 2}(\mathbf{x}_2, q_2)$.
	
	As a result, after this $i$-th iteration $u_2$ receives the messages $p^{i+1}_{2\leftarrow 1}(\mathbf{x}_1, q_1)$ and $p^{i+1}_{2\leftarrow 2}(\mathbf{x}_2, q_2)$ that do not recommend scheduling of $u_2$ at $c_1$ and $c_2$, respectively. 
	Therefore, $u_2$ requires its scheduling to be even stronger to allow both of the caching nodes to empty its queue, as the BP algorithm is iteratively repeated.
	In other words, $p^{i+1}_{2\rightarrow 1}(\mathbf{x}_1, q_1)$ and $p^{i+1}_{2\rightarrow 2}(\mathbf{x}_2, q_2)$ become considerably larger as $i$ increases.
	For this reason, both $c_1$ and $c_2$ make the final decision to schedule $u_2$ because of very large $p^I_{2\rightarrow 1}(x_{12} = 1)$ and $p^I_{2\rightarrow 2}(x_{22} = 1)$.
	Finally, two caching nodes intend to schedule $u_2$ at the same time; therefore, $\mathcal{I} ( \sum_{c_m \in \mathcal{H}_2} x_{m2} \leq 1 ) = 0$ and $f_2(\mathbf{x}_{\mathcal{H}_2}, \mathbf{q}_{\mathcal{H}_2} )=0$.
	This example shows that, although the case where multiple caching nodes deliver contents to the identical user by using the indicator function in the objective function \eqref{op3:opt} for all $u_n \in \mathcal{U}$ is prevented, the BP algorithm could result in many-to-one scheduling.
	Thus, using the BP algorithm alone may not achieve the maximum $F(\mathbf{x}, \mathbf{q})$.
	
	\subsection{Matching Algorithm for One-to-One Link Scheduling}
	\label{subsec:matching}
	
	As explained in Section \ref{subsec:prob_BP}, the user scheduling policy at every caching node based on \eqref{eq:decision_process} does not suffice to make each active user receive the content from only one node.
	To tackle this inherent problem of the BP algorithm, the one-to-one matching theory can be applied after the BP algorithm.
	We now define the matching $\Psi$, which indicates link scheduling between caching nodes and content-requesting users for content delivery in wireless caching networks.
	\begin{definition}
		A matching $\Psi$ is defined as
		\begin{align}
		&\Psi (c_m) \in \mathcal{U} \cup \{\emptyset \},~\forall c_m \in \mathcal{C} \label{eq:def-1} \\
		&\Psi^{-1} (u_n) \in \mathcal{C} \cup \{\emptyset \},~\forall u_n \in \mathcal{U} \label{eq:def-2} \\
		&|\Psi(c_m)| \in \{0, 1\},~\forall c_m \in \mathcal{C} \label{eq:def-3} \\
		&|\Psi^{-1}(u_n)| \in \{0, 1\},~\forall u_n \in \mathcal{U} \label{eq:def-4} \\
		&\Psi(c_m) = u_n \iff \Psi^{-1}(u_n) = c_m. \label{eq:def-5}
		\end{align}
	\end{definition}
	
	Specifically, $\Psi(c_m)$ indicates the user scheduled by $c_m \in \mathcal{C}$, and the scheduled user should be in the user set $\mathcal{U}$. 
	The case where $c_m$ is in the idle state is denoted by $\Psi(c_m) = \emptyset$ and $|\Psi(c_m)| = 0$; therefore, \eqref{eq:def-1} is satisfied.
	$\Psi^{-1}$ is the inverse function of $\Psi$, and $\Psi^{-1}(u_n)$ indicates the caching node determined for delivering to deliver the content to $u_n$, as given by \eqref{eq:def-2}.
	Similarly, if $u_n$ is not scheduled, then $\Psi^{-1}(u_n) = \emptyset$ and $|\Psi^{-1}(u_n)| = 0$.
	Since $\Psi$ allows one-to-one scheduling or unscheduled caching nodes and users, \eqref{eq:def-3} and \eqref{eq:def-4} hold.
	Finally, $\Psi(c_m) = u_n$ if and only if $\Psi^{-1}(u_n) = c_m$, representing that $c_m$ is determined for delivering the content to $u_n$.
	Note that, although the goal is to construct scheduling that matches one caching node to one requesting user, the matching $\Psi$ is not a strict one-to-one matching, because some caching nodes can be in the idle state and some users might not receive any content at a given time.
	In this subsection, $F(\mathbf{x},\mathbf{q},\Psi)$ is used to show the dependency of the utility function on the matching $\Psi$, rather than $F(\mathbf{x}, \mathbf{q})$.
	
	\begin{algorithm}[t!]
		\caption{Link scheduling algorithm for content delivery
			\label{algo:user_pairing}}
		\small
		\begin{algorithmic}[1]
			\State{Initialize $\Psi(c_m) \leftarrow \emptyset,~\forall c_m \in \mathcal{C}$.}
			\State{$F(\mathbf{x}, \mathbf{q}, \Psi) \leftarrow 0$}
			\For{ $\forall c_m \in \mathcal{C}$}
			{
				\State{$\mathcal{E} \leftarrow \emptyset$}
				\State{Compute $\mathcal{P}_{mnl},~\forall u_n \in \mathcal{V}_m, l\in\{1,\cdots,L \}$}
				\While{true}
				\State{$\mathcal{P}^*_m \leftarrow \underset{u_n \in \mathcal{V}_m \setminus \mathcal{E},~ l\in \{1, \cdots, L \}}{\max}~\mathcal{P}_{mnl}$}
				\State{$u_{n^*} \leftarrow \underset{u_n \in \mathcal{V}_m \setminus \mathcal{E}}{\arg\max}~ \underset{l\in\{1,\cdots,L \}}{\max}~ \mathcal{P}_{mnl}$}
				\If{$\mathcal{P}^*_m < \mathrm{Pr}\{\sum_{v\in \mathcal{V}_m} x_{mv} = 0 \}$}
				\State{$\Psi(c_m) \leftarrow \emptyset$}
				\State{break;}
				\Else
				\State{$\mathcal{E} \leftarrow \mathcal{E} \cup \{u_{n^*}\}$}
				\State{$\Psi' \leftarrow \textit{MatchRequest}(c_m, u_{n^*}, \Psi, \mathcal{E})$}
				\State{Compute $F(\mathbf{x}, \mathbf{q}, \Psi')$}
				\If{$F(\mathbf{x}, \mathbf{q}, \Psi') > F(\mathbf{x}, \mathbf{q}, \Psi)$}
				\State{$\Psi \leftarrow \Psi'$}
				\State{$F(\mathbf{x}, \mathbf{q}, \Psi) \leftarrow F(\mathbf{x}, \mathbf{q}, \Psi')$}
				\State{break;}
				\Else
				\State{$\mathcal{E} \leftarrow \mathcal{E} \cup \{u_{n^*} \}$}
				\EndIf
				\EndIf
				\EndWhile
			}
			\EndFor
		\end{algorithmic}
	\end{algorithm}
	
	Before the matching is found, matching values of every caching node and user are initialized; i.e., $\Psi(c_m) = \emptyset$ and $\Psi^{-1}(u_n) = \emptyset$ for all $c_m \in \mathcal{C}$ and $u_n \in \mathcal{U}$.
	Then, the matching $\Psi$ is constructed according to the marginalized distributions of link scheduling and power allocation at caching nodes $c_m$, i.e., $p(\mathbf{x}_m, q_m)$, for all $c_m  \in \mathcal{C}$.
	When finding the matching pair of $c_m$, $c_m$ prefers to schedule $u_n$ rather than to $u_k$ when the following inequality is satisfied:
	\begin{equation}
	\underset{l\in\{1,\cdots,L \}}{\max}~ \mathcal{P}_{mnl} > \underset{l\in\{1,\cdots,L \}}{\max}~ \mathcal{P}_{mkl}. \label{eq:prefer}
	\end{equation}
	According to constraint \eqref{eq:prefer}, $c_m$ can create a preference user list.
	Therefore, $c_m$ sends the matching request to the user in the most preferred order.
	Note that matching $\Psi$ determines link scheduling given the marginalized distributions $p(\mathbf{x}_m, q_m)$ for all $c_m\in \mathcal{C}$ obtained by the BP algorithm; therefore, when $\Psi(c_m) = u_n$ is determined, $c_m$ allocates the power level $P_{l^*}$ giving the largest marginalized probability as 
	\begin{equation}
	P_{l^*} = \underset{l\in\{1,\cdots,L \}}{\arg\max}~ \mathcal{P}_{mnl}.
	\end{equation} 
	
	Because excessive complexity is involved in computing and comparing marginalized probabilities for all possible link scheduling combinations, we focus on seeking a stable matching by using the deferred acceptance (DA) procedure \cite{Book:matching}. 
	Each caching node sends the matching request to the most preferred user, and the user who receives the request can accept or reject scheduling with the sender by comparing utility functions. 
	The link scheduling algorithm based on the matching theory is shown in Algorithm \ref{algo:user_pairing}, and details of the matching request and the decision for the received request are described in Algorithm \ref{algo:matching_request}.
	
	For example, suppose that $c_m$ sends the matching request to $u_n$ in the matching $\Psi$.
	Let $\Psi'$ be the optimal matching when $u_n$ accepts the request from $c_m$. 
	Then, $u_n$ decides whether to accept or reject the request from $c_m$ by comparing $F(\mathbf{x}, \mathbf{q}, \Psi)$ and $F(\mathbf{x}, \mathbf{q}, \Psi')$.
	$F(\mathbf{x}, \mathbf{q}, \Psi)$ is already obtained with the current matching $\Psi$, and $F(\mathbf{x}, \mathbf{q}, \Psi')$ must be computed. 
	When $\Psi^{-1}(u_n) = \emptyset$, the matching request is simply accepted when $f_n( \mathbf{x}_{\mathcal{H}_n}, \mathbf{q}_{\mathcal{H}_n} ) > 0$.
	Then, the optimal matching becomes $\Psi' = \Psi \setminus \{c_m, \emptyset \} \cup \{c_m, u_n \}$.
	
	However, when $\Psi^{-1}(u_n) \neq \emptyset$ if $u_n$ accepts the matching request from $c_m$, $\Psi^{-1}(u_n)$ should find another pair for constructing the optimal matching $\Psi'$. 
	According to Algorithm \ref{algo:matching_request}, $\Psi^{-1}(u_n)$ sends the matching request to its most preferred user, except for $u_n$. 
	If the most preferred user of $\Psi^{-1}(u_n)$ is $u_k$ and $\Psi^{-1}(u_k)=\emptyset$, then $\Psi' = \Psi \setminus \{(\Psi^{-1}(u_n), u_n) \} \cup \{ (\Psi^{-1}(u_n), u_k) \}$.
	Meanwhile, if $\Psi^{-1}(u_k) \neq \emptyset$, Algorithm \ref{algo:matching_request} should be recursively performed to construct $\Psi'$ until all the other nodes are matched. 
	Finally, $F(\mathbf{x}, \mathbf{q}, \Psi')$ is computed and compared with $F(\mathbf{x}, \mathbf{q}, \Psi)$.
	If $F(\mathbf{x}, \mathbf{q}, \Psi) < F(\mathbf{x}, \mathbf{q}, \Psi')$, the matching request from $c_m$ to $u_n$ is accepted, and $\Psi$ is updated by $\Psi'$, as shown in Algorithm \ref{algo:user_pairing}.
	
	\begin{algorithm}[t!]
		\caption{Matching request algorithm
			\label{algo:matching_request}}
		\small
		\begin{algorithmic}[1]
			\State{$\mathbf{Input}$: $c_m$, $u_n$, $\Psi$, and $\mathcal{E}$.}
			\State{$\mathbf{Output}$: $\Psi'$}
			\State{$\Psi' \leftarrow \Psi$}
			\State{$\Psi'(c_m) \leftarrow u_n$}
			\If{$\Psi^{-1}(u_n) \neq \emptyset$}
			\State{$c_k \leftarrow \Psi^{-1}(u_n)$}
			\If{$\mathcal{V}_k \subseteq \mathcal{E}$}
			\State{$\Psi'(c_k) \leftarrow \emptyset$}
			\Else
			\State{Compute $\mathcal{P}_{kjl},~\forall u_j \in \mathcal{V}_k, ~\forall l\in\{1,\cdots,L \}$}
			\State{$\mathcal{P}^*_{k} \leftarrow \underset{u_j \in \mathcal{V}_k \setminus \mathcal{E},~ l\in \{1,\cdots,L \}}{\max}~\mathcal{P}_{kjl}$}
			\State{$u_{j^*} \leftarrow \underset{u_j \in \mathcal{V}_k \setminus \mathcal{E}}{\arg\max}~ \underset{l\in\{1,\cdots,L \} }{\max}~\mathcal{P}_{kjl}$}
			\If{$\mathcal{P}^*_k < \mathrm{Pr}\{\sum_{v\in \mathcal{V}_k} x_{kv} = 0 \}$}
			\State{$\Psi'(c_k) \leftarrow \emptyset$}
			\Else
			\State{$\mathcal{E} \leftarrow \mathcal{E} \cup \{u_{j^*}\}$}
			\State{$\Psi' \leftarrow \textit{MatchRequest}(c_k, u_{j^*}, \Psi', \mathcal{E})$}
			\EndIf
			\EndIf
			\EndIf
		\end{algorithmic}
	\end{algorithm}
	
	The optimal link scheduling based on $p(\mathbf{x}_m, q_m)$ for all $c_m \in \mathcal{C}$ can be obtained by searching all possible combinations of link scheduling to find $\Psi$ that maximizes $F(\mathbf{x}, \mathbf{q}, \Psi)$. 
	Then, the system needs to search $\prod_{m=1}^{M} |\mathcal{V}_m|$ combinations exhaustively to find the optimal link scheduling, and the time complexity is approximately $O(N^M)$, assuming that $N \approx |\mathcal{V}_m|$.
	In the proposed matching algorithm, the worst case is that every caching node can generate the signal link with all $N$ users and $c_m$ undergoes a trial and error operation when sending the matching request to users already matched with $c_1,\cdots, c_{m-1}$.
	This requires $\sum_{m=1}^M m = \frac{m(m+1)}{2}$ comparison steps, and the time complexity becomes approximately $O(M^2)$, which is considerably smaller than $O(N^M)$, because $N>M$ and $M\geq2$ in general.
	Note that the complexity gain of the proposed algorithm grows as $N$ increases and the worst case of the proposed matching algorithm rarely occurs.
	
	\section{Numerical results}
	\label{sec:numerical_results}
	
	In this section, we show that the proposed algorithm for distributed and dynamic link scheduling and power allocation is effective in the caching helper network, as well as in the D2D-assisted caching network. 
	The simulation parameters listed in Table \ref{table:parameters} are used unless otherwise noted.
	
	\begin{table}[t!]
		\small
		\caption{System Parameters for Simulation}
		\label{table:parameters}
		\begin{center}
			\scalebox{1}{
				\begin{tabular}{|l|c|l|c|}
					\hline 
					Bandwidth ($\mathcal{B}$) & 10 MHz & Max. power budget ($q_{\text{max}}$) & 2 W \\
					\hline
					Chunk size ($S$) & 20 kbits & Coherence time ($\tau_c$) & 10 ms \\
					\hline
					Radius of signal link coverage ($d_s$) & 100 & Radius of interference link coverage ($d_i$) & 300 \\
					\hline
					Path loss exponent ($\alpha$) & 3 & Noise variance ($\sigma^2$) & $10^{-8}$ \\
					\hline
				\end{tabular}
			}
		\end{center}
		\vspace{-5mm}
	\end{table}
	\vspace{-2mm}
	
	\subsection{Simulation Environments}
	\label{subsec:simul_env}
	
	The simulations assume a frequency-flat Rayleigh fading channel, and the amplitude gain between $c_m$ and $u_n$ is denoted by $h_{m,n}(t) = \sqrt{D_{m,n}}g(t)$, where $D_{mn} = 1/d_{mn}^{\alpha}$ denotes path gain (the inverse of path loss). 
	In addition, $d_{mn}$ and $\alpha$ are the distance between $c_m$ and $u_n$ and the path loss exponent, respectively. 
	$g(t)$ represents the fast fading component at slot $t$ having a circularly symmetric complex Gaussian distribution, $g(t) \sim \mathcal{CN}(0,1)$.
	
	The simulation results were obtained for two different scenarios: 1) the caching helper network (Fig. \ref{fig:network_model}) and 2) the D2D-assisted caching network (Fig. \ref{fig:d2d_model}).
	In Fig. \ref{fig:network_model}, there are three caching helpers each having a coverage region with radius $R$, and the locations of the helpers are $[(0,0),(\frac{5}{3}R, 0), (\frac{5}{6}R, \frac{5\sqrt{3}}{6}R)]$, and thus, their coverage regions are partially overlapped.
	Users are randomly distributed according to a homogeneous PPP with intensity $\lambda = 0.01 \times 10^{-2}$.
	Here, if the caching helper stores contents requested by users in its coverage region, they can generate the signal link; i.e., $d_s = R$. 
	In addition, $d_i = 3R$; therefore, the caching helper can interfere with users outside its coverage.
	
	In the caching helper network, the proposed link scheduling and power allocation policy was verified by comparison with the optimal scheme that is obtained by the exhaustive search technique.
	Since the complexity required to search exhaustively the optimal link scheduling and power allocation significantly increases as the number of caching nodes grows, this comparison operation is demonstrated in a caching helper network with only three helpers. 
	Furthermore, we demonstrate the advantage of using the matching algorithm after the BP algorithm has been implemented. 
	Thus, the three techniques we compared are as follows.
	\begin{itemize}
		\item ``Exhaustive search": This scheme exhaustively searches the link scheduling and power allocation that minimize the upper bound on the drift-plus-penalty term in \eqref{eq:dpp_upper}.
		It can be considered as the numerically optimal scheme and a centralized decision strategy.
		
		\item ``BP with matching": This scheme is the proposed content delivery strategy. 
		The marginalized probability distributions of link scheduling and power allocation at every caching node are estimated by using the BP algorithm. 
		Then, the matching algorithm helps to construct the one-to-one scheduling network.
		
		\item ``BP without matching": The proposed BP-based link scheduling and power allocation are employed in this technique, but the matching algorithm is not used. 
	\end{itemize}
	
	In the D2D-assisted caching network (Fig. \ref{fig:d2d_model}),
	user devices are randomly distributed in the square region of 600 m x 600 m based on an independent PPP with intensity $\lambda = 0.04 \times 10^{-2}$.
	All the devices can cache some popular contents. 
	Each device can request the content with the activity probability $p_a = 0.2$.
	All the other inactive devices can act as caching nodes that can deliver the cached content to the nearby content-requesting user.
	Therefore, the number of active users is considerably smaller than the number of potential caching nodes, and this situation is consistent with the practical D2D scenario. 
	Some content-requesting users may not find any caching node for receiving the desired content, and we assumed that they receive the desired content from the BS or the server; therefore, these users were ignored in the simulation. 
	$d_s = 100$ and $d_i = 300$ are used as in the caching helper.
	Because there are so many potential caching nodes, the complexity of the BP algorithm is very large; therefore, the approximated BP algorithm was used.
	In large-scale D2D-assisted caching networks, clustering and spatial reuse constitute a popular scheduling technique (see Section \ref{sec:intro}). 
	Thus, in the simulations the following approaches were compared.
	\begin{itemize}
		\item ``Approximated BP with matching": 
		When computing $p^i_{n \rightarrow m} (\mathbf{x}_m, q_m)$, the operation of averaging $p^i_{n \leftarrow m} ( \mathbf{x}_k, q_k )$ for all $c_k \in \mathcal{H}_n \setminus \{c_m \}$ requires excessive complexity when the number of caching nodes is large. 
		To reduce its computations, we choose a small subset of $\mathcal{H}_n$, which is denoted by $\mathcal{N}_n \subseteq \mathcal{H}_n$. 
		Then, $p^i_{n \rightarrow m}(\mathbf{x}_m, q_m)$ is computed by averaging 
		$p^i_{n \leftarrow m} ( \mathbf{x}_k, q_k )$ for all $c_k \in \mathcal{N}_n \setminus \{c_m \}$ and assuming that other caching nodes, i.e., $c_v \in \mathcal{H}_n \setminus \mathcal{N}_n$, are consuming their expected power $\bar{q}_v = \sum_{l=1}^L \mathrm{Pr}\{q_v = P_l \} \cdot P_l$.
		Therefore, the approximated factor node update becomes
		\begin{equation}
		p^i_{n \rightarrow m}(\mathbf{x}_m, q_m) = \mathbb{E} \Big[ \exp \Big( \delta f_n( \mathbf{x}_{\mathcal{H}_n}, \mathbf{q}_{\mathcal{H}_n} ) \Big) \Big| \mathbf{x}_m, q_m, \bar{q}_v, \forall v \in \mathcal{N}_n \Big]. \label{eq:approximated_BP}
		\end{equation}
		In this simulation, only the nearest caching node to $u_n$ in $\mathcal{H}_n$ was selected for $\mathcal{N}_n$; i.e., $|\mathcal{N}_n| = 1$. 
		
		\item ``Clustering method 1": 
		Since devices can interfere with each other, the entire region is divided into nine clusters in a grid manner, as shown in Fig. \ref{fig:d2d_model}.
		Each cluster is square-shaped, and the width is 200 m. 
		As assumed in most existing caching studies supposed, this method allows only one active delivery link within a cluster to remove the intra-cluster interference. 
		In addition, because $d_i = 300$, the delivery link in a specific cluster can interfere with all active users in the eight adjacent clusters; therefore, we assume that different clusters access exclusive frequency bands with the equally divided bandwidth of $\frac{\mathcal{B}}{9}$.
		
		\item ``Clustering method 2":
		The structures of clusters and use of orthogonal bands are the same as in ``Clustering method 1". 
		Meanwhile, the proposed delivery scheme is employed in each cluster separately.
		Thus, there is no inter-cluster interference as a result of using orthogonal bands, and the proposed technique manages intra-cluster interference within each cluster. 
		
		\item \cite{TCOM2016Zhang}:
		The D2D link scheduling and power allocation scheme in \cite{TCOM2016Zhang} is compared with the proposed delivery policy. 
		The scheme in \cite{TCOM2016Zhang} maximizes sum-throughput and it is a centralized decision made with comprehensive knowledge of channel information. 
	\end{itemize}
	
	In both simulation cases, i.e., the helper network and the D2D-assisted network, we assumed that the content placements in the caching nodes were already completed. 
	Taking into account wireless fading channels and stochastic geometry for user distributions, the probabilistic caching policy proposed in \cite{TWC2016Chae} was used. 
	
	\begin{figure}[t]
		\minipage{0.37\textwidth}
		\includegraphics[width=\linewidth]{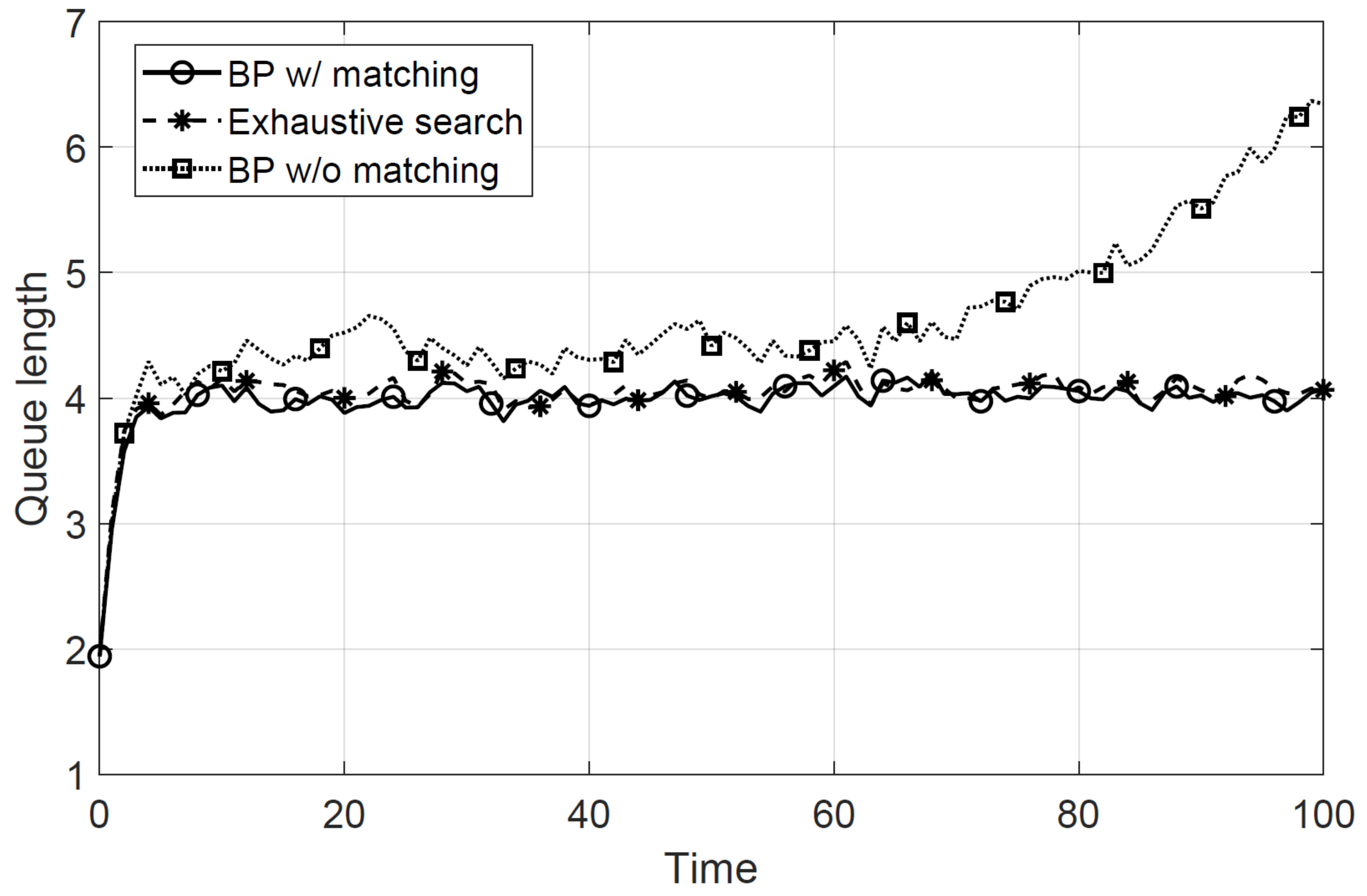}
		\caption{Average queue lengths in the caching helper network}
		\label{fig:queue_helper}
		\endminipage\hfill
		\minipage{0.40\textwidth}
		\includegraphics[width=\linewidth]{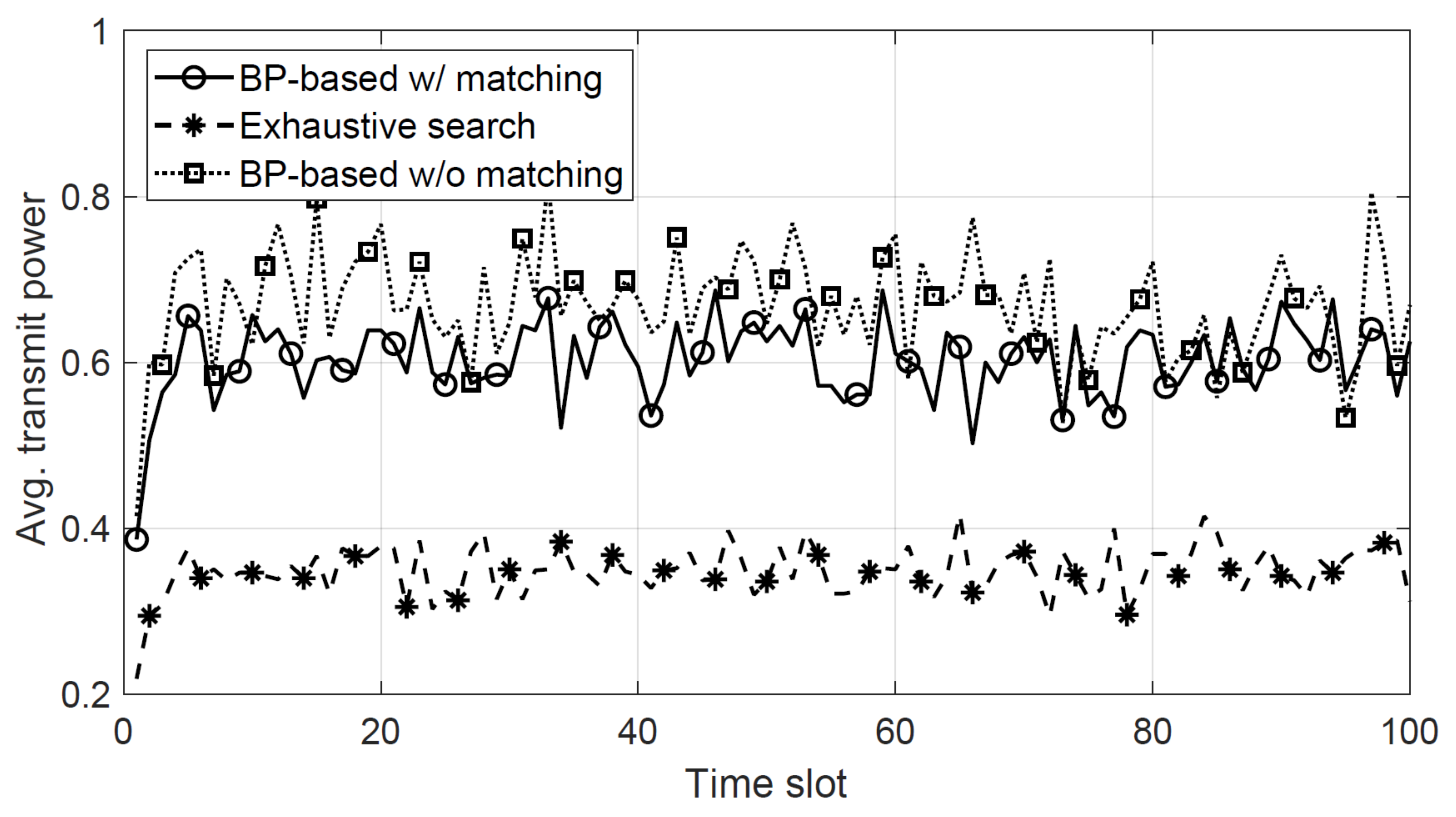}
		\caption{Average power sum in the caching helper network}
		\label{fig:power_avg}
		\endminipage
		\vspace{-3mm}
	\end{figure}
	
	\subsection{Numerical Results for Caching Helper Networks}
	\label{subsec:helper_network}
	
	The average queue length (i.e., user demands) and the average power consumption per each time slot in the caching helper network are shown in Figs. \ref{fig:queue_helper} and \ref{fig:power_avg}, respectively.
	We can see that the average queue length and power consumption of the proposed scheme and ``Exhaustive search" are clearly upper bounded, in accordance with Lyapunov theory. 
	In fact, the tightness of these upper bounds is not explicitly ensured by the min-drift-plus-penalty algorithm; nevertheless, the average queue length of the proposed scheme is almost the same as that of ``Exhaustive search."
	Therefore, the proposed scheme can provide an average queueing delay that is almost the same as the optimal one, at the expense of a 70$\%$ increase in transmit power. 
	However, if the matching algorithm is not employed, the average queue length increases after some time. 
	
\begin{figure}[t]
		\minipage{0.3\textwidth}
		\includegraphics[width=\linewidth]{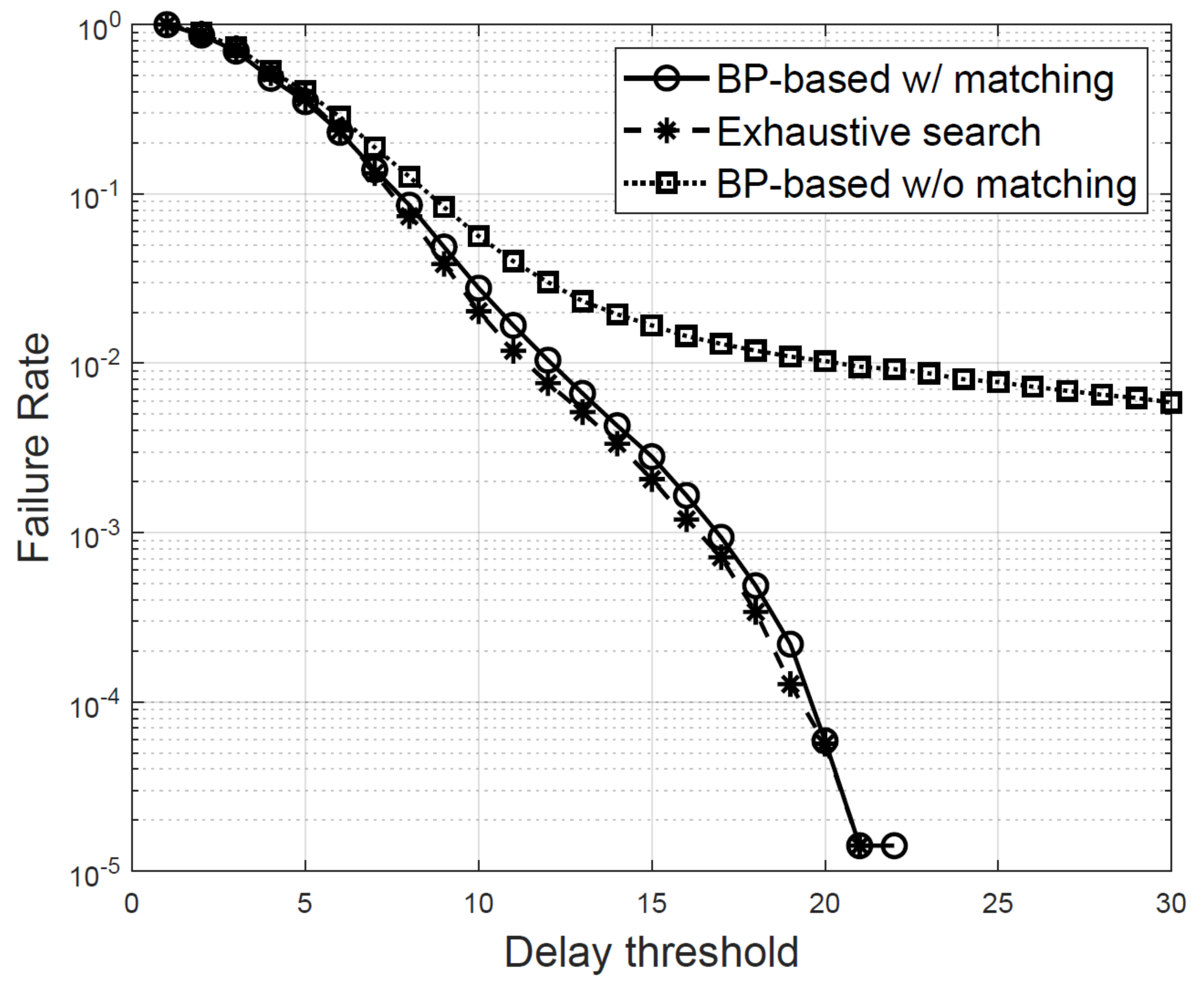}
		\caption{Failure rates in the caching helper network}
		\label{fig:failure_helper}
		\endminipage\hfill
		\minipage{0.3\textwidth}
		\includegraphics[width=\linewidth]{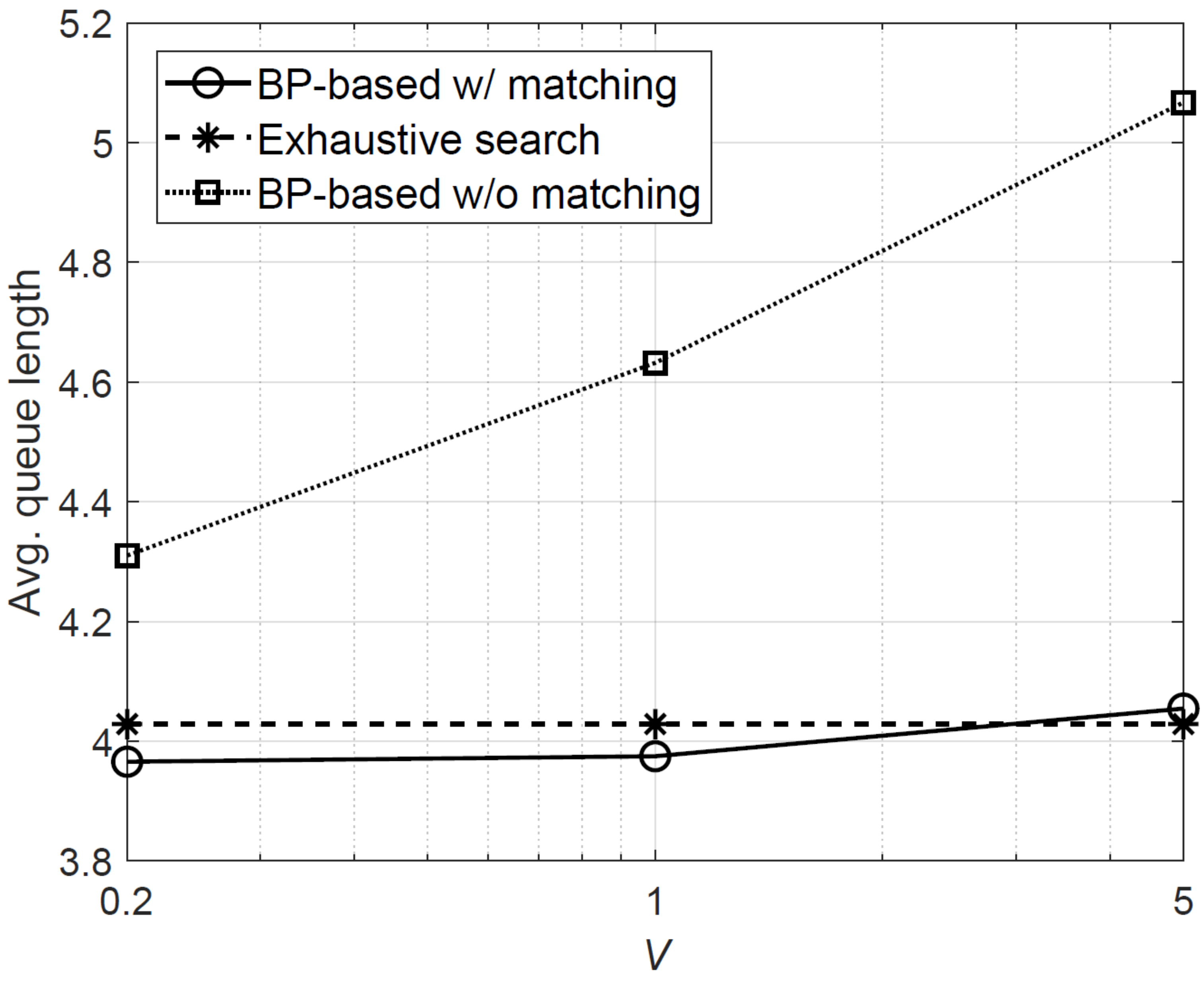}
		\caption{Effect of $V$ on average queue length}
		\label{fig:avg_queue_Vlist}
		\endminipage\hfill
		\minipage{0.3\textwidth}
		\includegraphics[width=\linewidth]{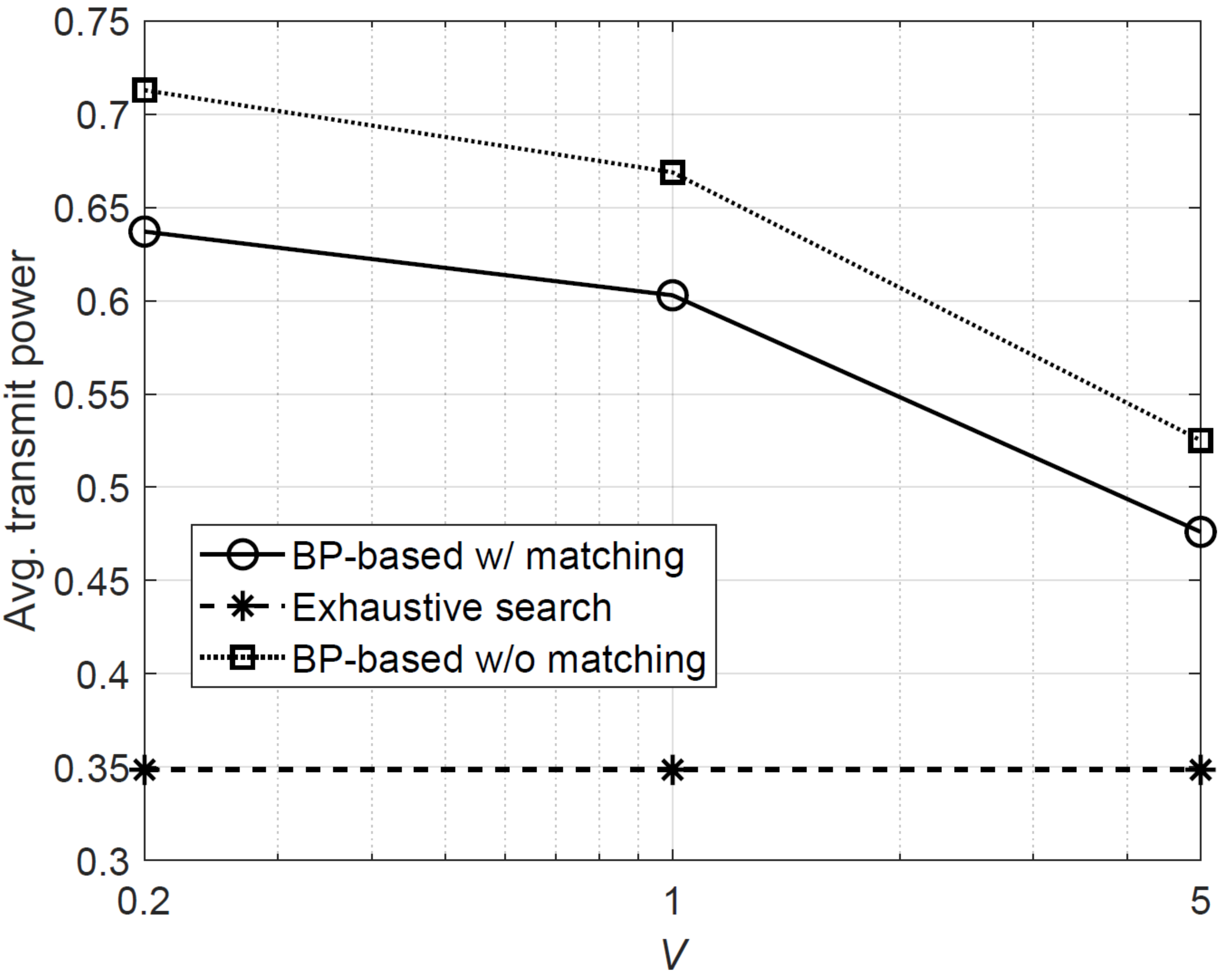}
		\caption{Effect of $V$ on average transmit power}
		\label{fig:power_Vlist}
		\endminipage\hfill
	\end{figure}

	In addition, the instantaneous delay is as important as the average queuing delay.
	When users impose a stringent constraint on the delay performance, a service failure can be defined as the event that user demands are provided after the predetermined delay threshold.
	In this scenario, where all the requested contents have an identical delay threshold, service failure rates can be obtained as shown in Fig. \ref{fig:failure_helper}.
	In this figure, the proposed scheme shows almost the same failure rate as ``Exhaustive search"; however, ``BP without matching" yields a considerably higher failure rate than the other techniques.
	
	In Figs. \ref{fig:avg_queue_Vlist} and \ref{fig:power_Vlist}, the effect of the system parameter $V$ on the average queue length and power consumption, respectively, is observed.
	As shown in \eqref{eq:dpp_upper}, $V$ is an importance factor for power efficiency; therefore, as $V$ increases, power consumption decreases and the average queue length grows. 
	Because ``Exhaustive search" is the numerically optimal technique, its performance does not vary considerably with $V$; however, the transmit power sums and average queue lengths of the BP-based techniques decrease and increase, respectively. 
	In Figs. \ref{fig:queue_helper} and \ref{fig:power_avg} obtained with $V=1$, we show that ``BP with matching" consumes 70$\%$ more power than ``Exhaustive search" to provide almost the same queueing delay.
	However, Figs. \ref{fig:avg_queue_Vlist} and \ref{fig:power_Vlist} show that, if larger $V=5$ is used, ``BP with matching" can save significant power (i.e., more than 35$\%$ of the power compared to ``Exhaustive search") while limiting the queueing delay on a scale similar that of ``Exhaustive search".
	Note that $V$ is a system parameter, the appropriate value of which must be found empirically. It can also be adjusted freely by the system designer. 
	
	\subsection{Numerical Results for Device-to-Device-Assisted Caching Network}
	\label{subsec:d2d_network}

	In the D2D-assisted caching network, the number and locations of both the devices that deliver and the devices that receive contents are not fixed, and potential transmitters do not have their own coverage regions. 
	This results in a higher probability that users experience interference from multiple caching nodes than users in the helper network. 
	However, the benefit for users is that there are many nearby potential caching nodes that can deliver the content. 
	
	\begin{figure}[t]
		\minipage{0.37\textwidth}
		\includegraphics[width=\linewidth]{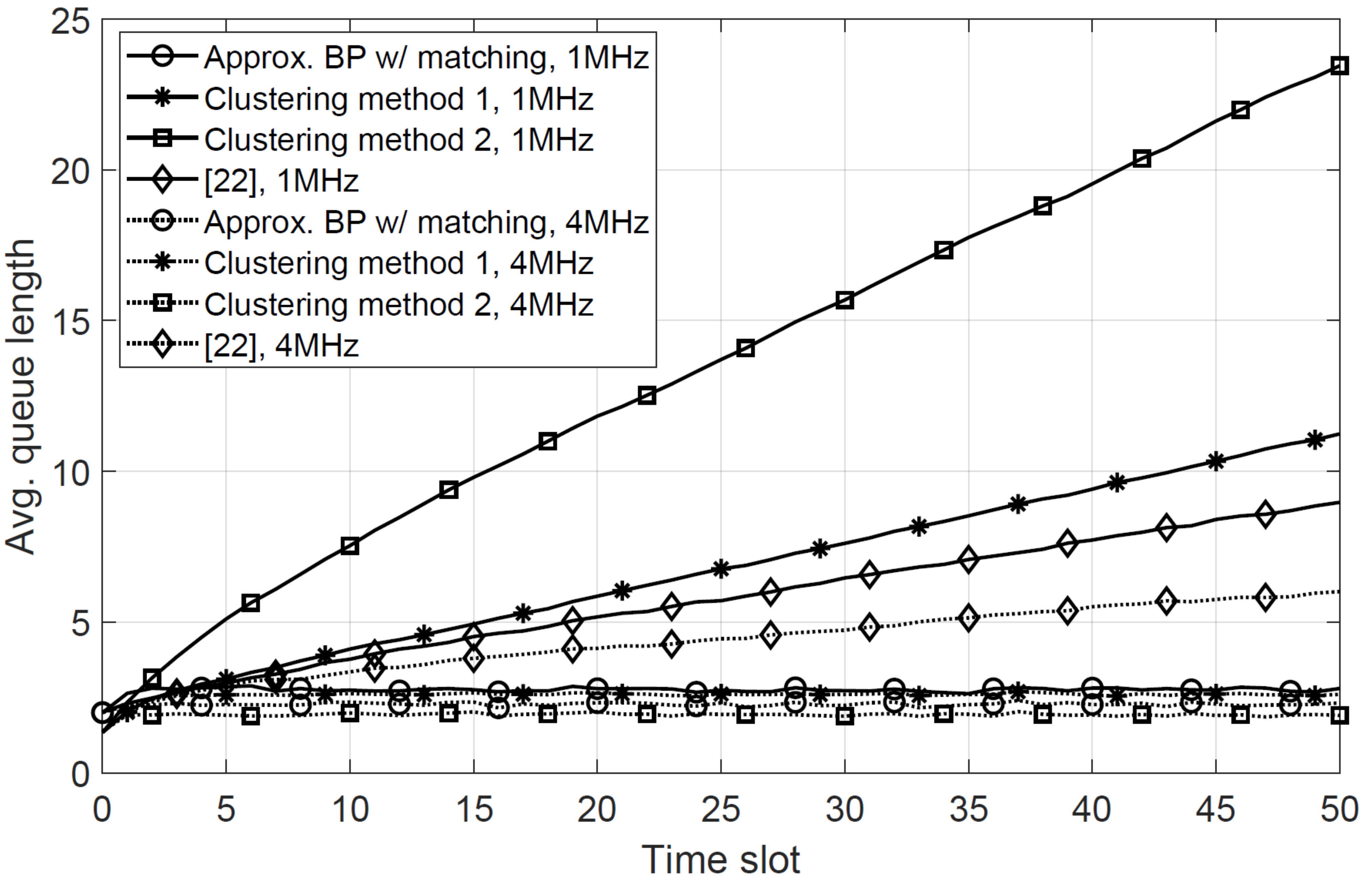}
		\caption{Averaged queue lengths in the D2D-assisted caching network}
		\label{fig:queue_d2d}
		\endminipage\hfill
		\minipage{0.37\textwidth}
		\includegraphics[width=\linewidth]{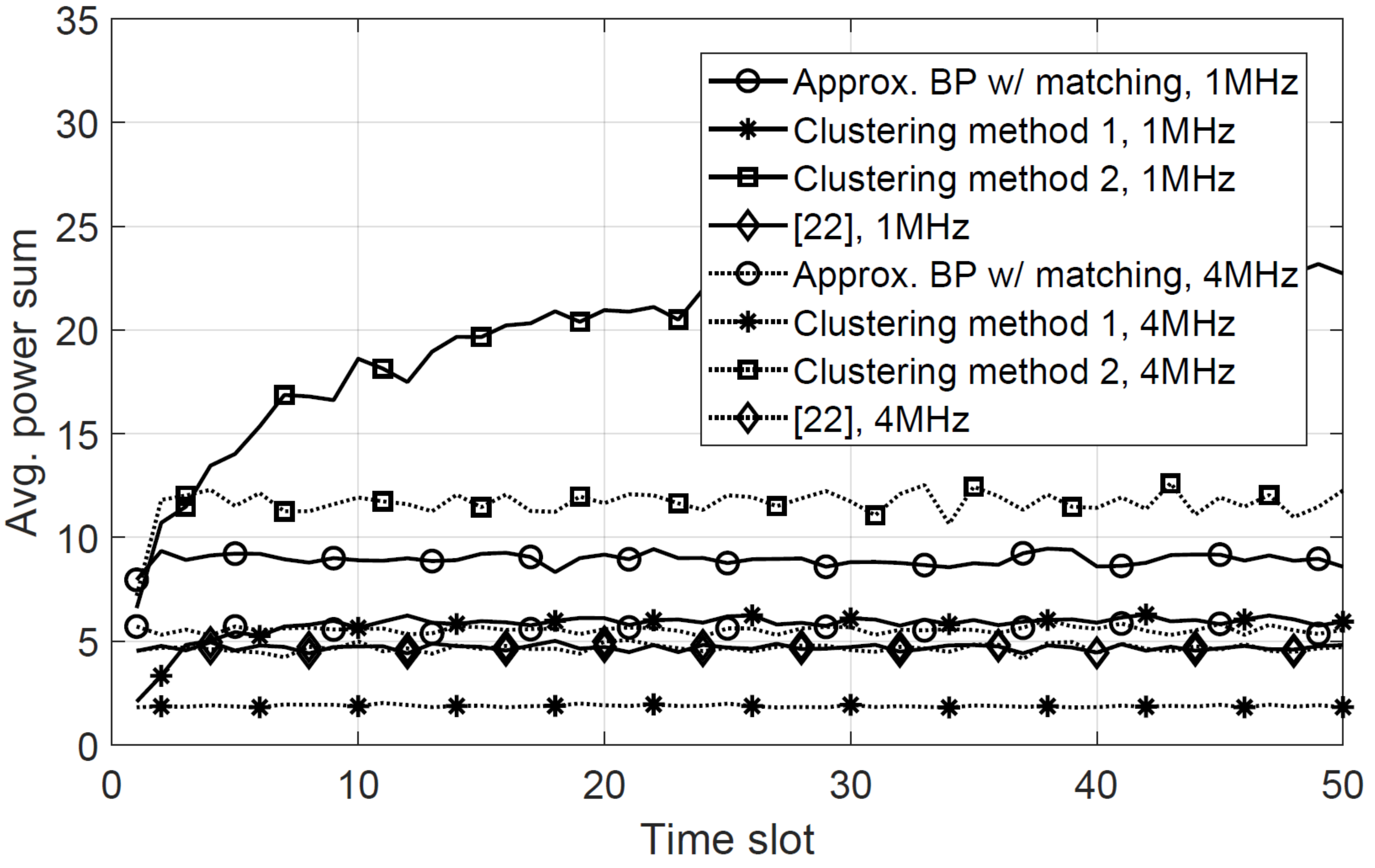}
		\caption{Average power sum in the D2D-assisted caching network}
		\label{fig:power_d2d}
		\endminipage
		\vspace{-3mm}
	\end{figure}
	
	\begin{figure}[t]
		\minipage{0.37\textwidth}
		\includegraphics[width=\linewidth]{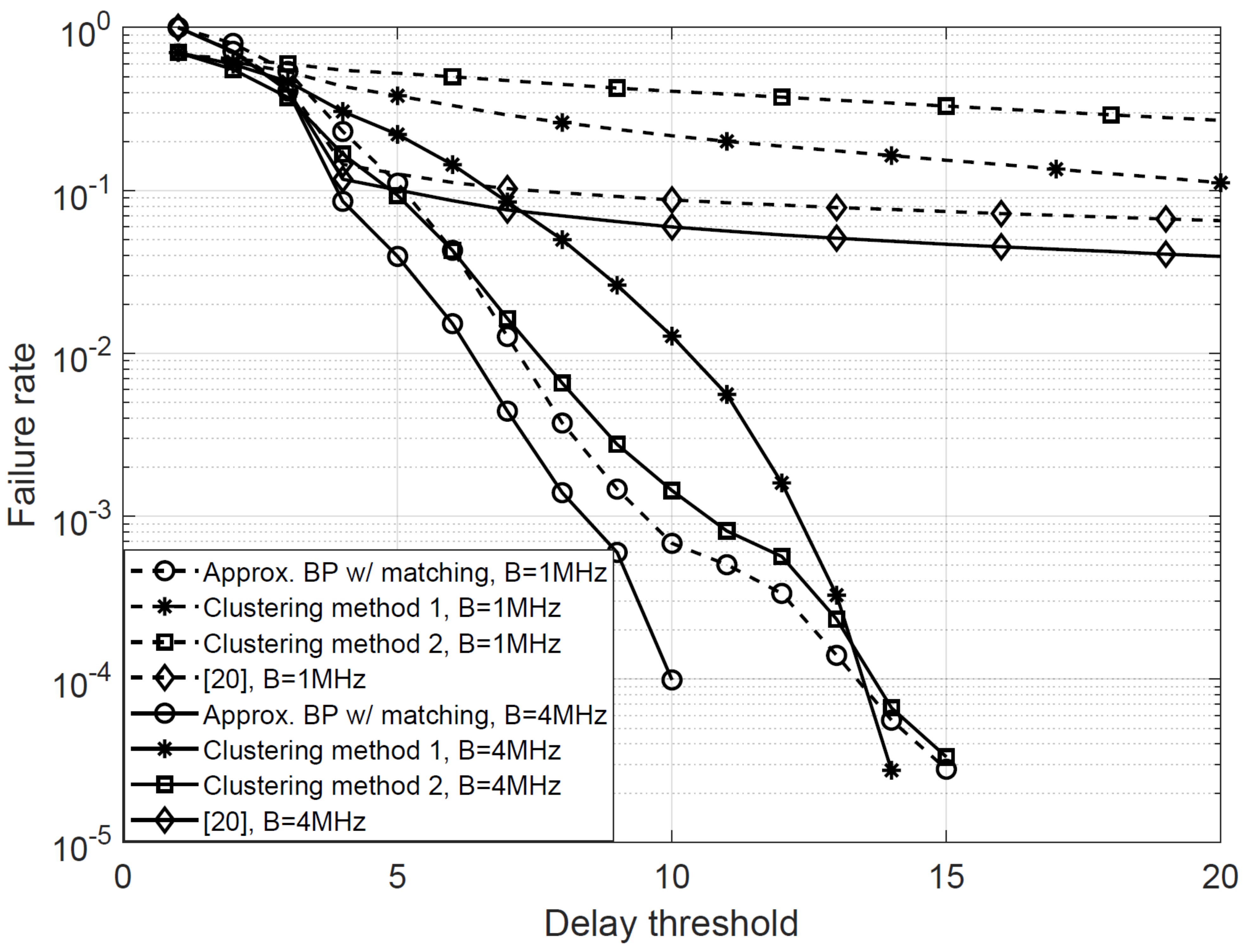}
		\caption{Failure rates in the D2D-assisted caching network}
		\label{fig:failure_d2d}
		\endminipage\hfill
		\minipage{0.39\textwidth}
		\includegraphics[width=\linewidth]{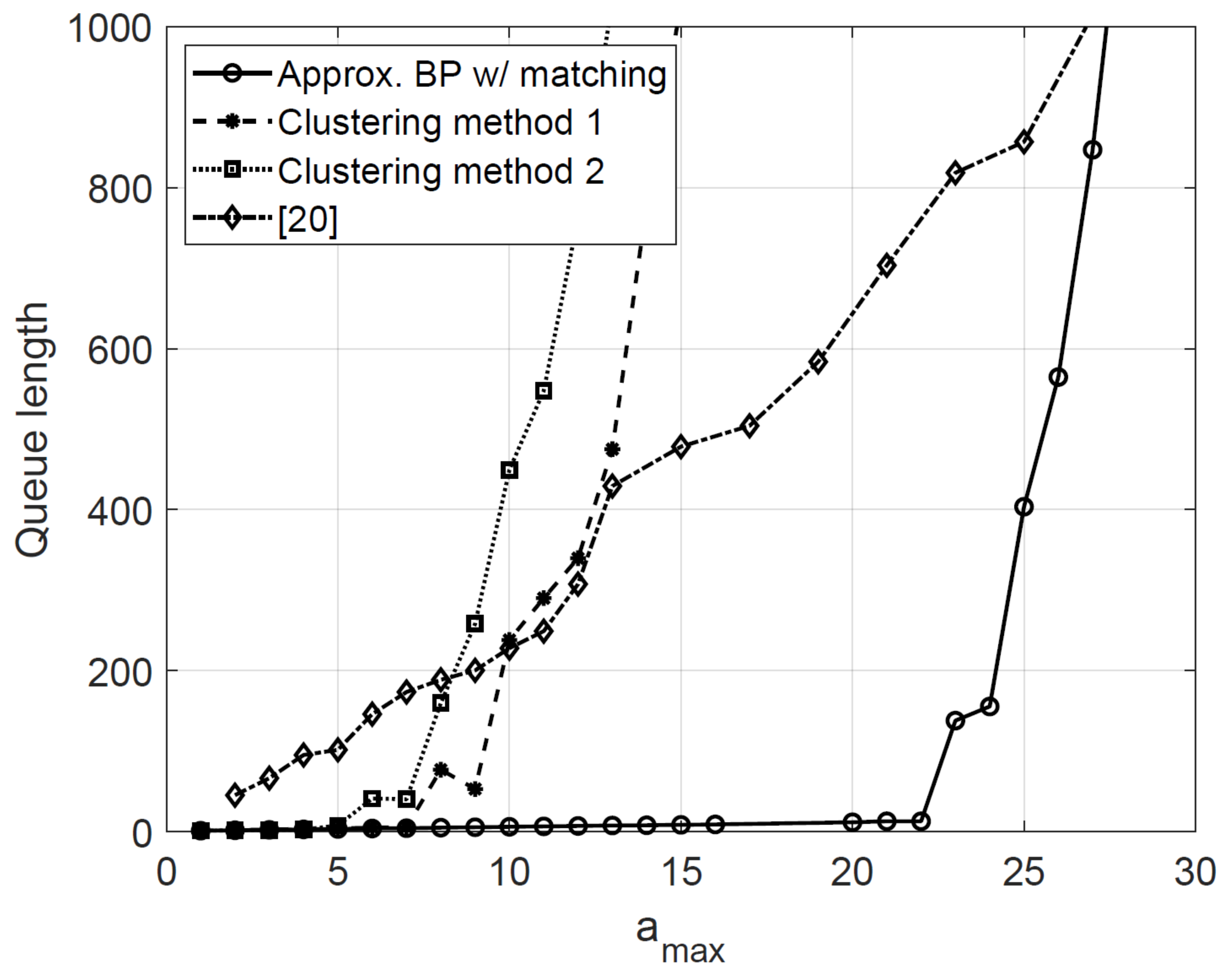}
		\caption{Queue length versus $a_{\text{max}}$}
		\label{fig:queue_a_list}
		\endminipage
		\vspace{-3mm}
	\end{figure}

	Fig. \ref{fig:queue_d2d} shows time-averaged queue lengths versus time slots with $\mathcal{B}=1$ MHz and $\mathcal{B}=4$ MHz.
	If a sufficiently large bandwidth is available, i.e., $\mathcal{B}=4$ MHz, clustering methods can make the averaged queue lengths upper bounded; however, when $\mathcal{B}=1$ MHz clustering methods cannot satisfy user demands within a reasonable delay time. 
	However, the time-averaged queue length obtained by the proposed scheme using the approximated BP algorithm is upper bounded even when $\mathcal{B} = 1$ MHz. 
	Even when the accessible bandwidth becomes large, the converged queue length of the proposed scheme is similar to those of clustering methods. Therefore, the proposed delivery policy provides higher spectral efficiency than clustering methods and is useful over a wide range of frequency bandwidths even without any clustering.
	
	Meanwhile, the average power of ``Clustering method 1" is smaller than that of the proposed scheme and ``Clustering method 2" in Fig. \ref{fig:power_d2d}. 
	However, different comparison techniques have different numbers of caching nodes delivering contents to active devices; therefore, ``Clustering method 1" saves power consumption, because it does not manage interference effectively and many caching devices are in the idle state, especially when $\mathcal{B} = 1$ MHz.
	Thus, we note that the boundedness of power consumption is important herein rather than the absolute power level. 
	
	Here, it should be also noted that a comparison of the proposed technique and the scheme proposed in \cite{TCOM2016Zhang} shows that they differ in terms of many characteristics. 
	First, the goal of the scheme proposed in \cite{TCOM2016Zhang} is to maximize throughput rather than limit the queue length; therefore, the average queue length of \cite{TCOM2016Zhang} is not upper bounded, as shown in Fig. \ref{fig:queue_d2d}. 
	Specifically, throughput maximization could neglect the support of some users whose link rates are not large, i.e., user fairness is not captured effectively.
	Second, the scheme in \cite{TCOM2016Zhang} considers only one potential helper for each content-requesting user; i.e., link scheduling is considered but not node association. 
	However, because multiple inactive devices are potential transmitters in the proposed scheme, it is very advantageous to allow more caching nodes to deliver contents and thus satisfy more users' demands.
	Third, the link scheduling and power allocation proposed in \cite{TCOM2016Zhang} are determined in a centralized manner with comprehensive knowledge of network topology and channel information. 
	However, in the proposed scheme, distributed decisions on link scheduling and power allocation are made at the caching node side.
	Nevertheless, the proposed distributed scheme is superior to that in \cite{TCOM2016Zhang} in terms of supporting random user demands, at the expense of increased power consumption. 
	
	In Fig. \ref{fig:failure_d2d}, service failure rates versus the stringent delay constraint are shown. 
	With the average queue lengths shown in Fig. \ref{fig:queue_d2d}, the proposed scheme can support delay-sensitive communications considerably better than other comparison techniques, even when the bandwidth is not large, i.e., $\mathcal{B}=1$ MHz. 
	On the other hand, because the scheme proposed in \cite{TCOM2016Zhang} does not limit the queue backlog, its failure rate shows an error floor. 
	We can also see in Fig. \ref{fig:queue_a_list} that the proposed scheme is also robust to increasing the number of random user demands.
	The maximum value of $a_{\mathrm{max}}$, with which the queue backlog does not diverge, can be interpreted as the delivery link capacity to support multiple users in D2D-assisted caching networks. 
	Accordingly, Fig. \ref{fig:queue_a_list} shows that the proposed link scheduling and power allocation scheme provides fundamentally a considerably larger delivery link capacity than other comparison techniques.
	
	\section{Concluding Remarks}
	\label{sec:conclusion}
	
	In this paper, a scheme for distributed and dynamic link scheduling and power allocation for content delivery in wireless caching networks was proposed. 
	First, the joint optimization problem of link scheduling and power allocation was formulated based on a Lyapunov optimization framework.
	Then, the distributed decision process at each caching node depending on the probability distributions of link scheduling and power allocation was presented.
	The probability distributions of link scheduling and power allocation can be obtained by constructing the factor graph representing the network topology and using the BP-based link scheduling and power allocation scheme.
	The intrinsic problem of the proposed BP algorithm was shown, namely, that, although each user is allowed to be served by only one caching node owing to asynchronous demands, the BP algorithm frequently results in many-to-one scheduling.
	Therefore, a matching algorithm for one-to-one scheduling was also proposed to tackle this problem.
	The numerical results show that the proposed scheme enables users to pursue high power efficiency while the average queueing delay is limited.
	Although the proposed technique is based on distributed decisions, it provides a performance similar to that of the optimal scheme with comprehensive knowledge of the entire network.
	We showed that in the large-scale D2D network, the proposed scheme does not require a clustering method and/or use of orthogonal frequency bands and provides a fundamentally larger delivery link capacity than existing techniques.
	

	\vspace{12pt}

\end{document}